\documentclass[lettersize,journal]{IEEEtran}
\usepackage{amsmath,amsfonts}
\usepackage{array}
\usepackage[caption=false,font=normalsize,labelfont=rm,textfont=rm]{subfig}
\usepackage{textcomp}
\usepackage{stfloats}
\usepackage{url}
\usepackage{verbatim}
\usepackage{graphicx}
\usepackage{cite}
\usepackage{mathtools}
\usepackage{siunitx}
\usepackage{booktabs}
\RequirePackage[linesnumbered, ruled, vlined]{algorithm2e}
\usepackage{soul}

\usepackage{tikz, pgfplots}
\pgfplotsset{compat=1.18}
\usetikzlibrary{positioning,calc}
\usepackage{xcolor}
\usepackage{tabularx}
\usepackage{multirow}
\newcolumntype{Y}{>{\centering\arraybackslash}X}

\SetCommentSty{mycommfont}
\def\BibTeX{{\rm B\kern-.05em{\sc i\kern-.025em b}\kern-.08em
    T\kern-.1667em\lower.7ex\hbox{E}\kern-.125emX}}
\usepackage{balance}
\usepackage{hyperref}
\hypersetup{colorlinks=true, urlcolor=blue} 
\usepackage[capitalize]{cleveref}

\begin{document}

\title{A Distributionally Robust Multi-agent Reinforcement Learning Framework for Intelligent Intersection Control}
\author{Shuwei Pei, Joran Borger, Arda Kosay, Bayu Jayawardhana, Muhammed O. Sayin, Saeed Ahmed
\thanks{This work was supported by the Holland High Tech (TKI HTSM) strategic program PPS-I Flex HighTech under the project number 24PPS173-CABS.\\ Shuwei Pei, Joran Borger, Bayu Jayawardhana, and Saeed Ahmed are with Engineering and Technology Institute Groningen, Faculty of Science and Engineering, University of Groningen, 9747 AG Groningen, the Netherlands. Arda Kosay and Muhammed O. Sayin are with the Department of Electrical, Electronics Engineering, Bilkent University, TR-06800 Ankara, Turkey. Email: s.pei@rug.nl, j.borger.3@student.rug.nl, arda.kosay@bilkent.edu.tr, b.jayawardhana@rug.nl, sayin@ee.bilkent.edu.tr, s.ahmed@rug.nl}}

\maketitle

\begin{abstract}
Multi-agent reinforcement learning (MARL) has emerged as a promising approach for traffic signal control. However, standard MARL policies typically optimize for expected returns under nominal conditions, leaving them highly vulnerable to spatial-temporal demand shifts and catastrophic congestion under adverse scenarios. To address this critical limitation, this paper proposes an algorithm-agnostic Distributionally Robust (DR) MARL framework integrating an adaptive Contextual-Bandit Worst-Case Estimator (CB-WCE). Operating on a slower timescale, the CB-WCE co-evolves with the traffic controllers by dynamically generating adversarial demand mixtures during training. This steers the learning process to fortify policies against bottleneck scenarios without requiring modifications to the underlying MARL architectures. The framework is evaluated across value-based, actor-critic, and policy-gradient methods on both a synthetic $5\times 5$ grid and a heterogeneous Monaco City network. Empirical results demonstrate that the DR framework prevents unbounded queue growth and profoundly enhances both worst-case robustness and average-case efficiency. Notably, for the Proximal Policy Optimization (PPO) architecture in the Monaco environment, on average, robust retraining reduced the worst-case queue length by 74.39\% and improved the average-case network-wide queue length by 75.45\%. Furthermore, the retrained policies exhibit strong zero-shot generalization to unseen traffic distributions, highlighting the framework's scalability and potential for resilient real-world urban deployment.
\end{abstract}


\section{Introduction}

Signalized intersections are critical bottlenecks in urban road networks, contributing significantly to travel delays, fuel consumption, and pollutant emissions. Suboptimal signal timing exacerbates congestion, causing economic and environmental impacts~\cite{barth2008real}, while also worsening air quality and associated public health risks~\cite{stanaway2018global}. As urbanization and population growth increasingly strain saturated transport infrastructure~\cite{dijkstra2021applying}, developing traffic signal control strategies resilient to highly variable and uncertain demand remains a paramount challenge for sustainable urban mobility.

Conventional traffic signal control predominantly relies on fixed-time plans, actuated logic, or rule-based adaptive schemes. Widely deployed systems, such as SCOOT~\cite{Hunt1982SCOOT} and SCATS~\cite{Luk1984SCATSCOOT}, adapt cycle lengths and splits using historical data and localized real-time measurements. Although optimization-based approaches like OPAC~\cite{gartner1982demand} and PRODYN~\cite{henry1984prodyn} formulate signal control as a dynamic programming problem, their computational complexity hinders network-level scalability. Crucially, these traditional and heuristic methods degrade when traffic patterns deviate from nominal assumptions due to incidents, special events, or spatial imbalances. Coping with the inherent stochasticity of urban traffic is notoriously difficult; in dense networks, performance further deteriorates due to complex interactions involving queue spillback and blocking~\cite{stevanovic2010atcs}.
\begin{figure}[t]
    \centering
    \includegraphics[width=1\linewidth]{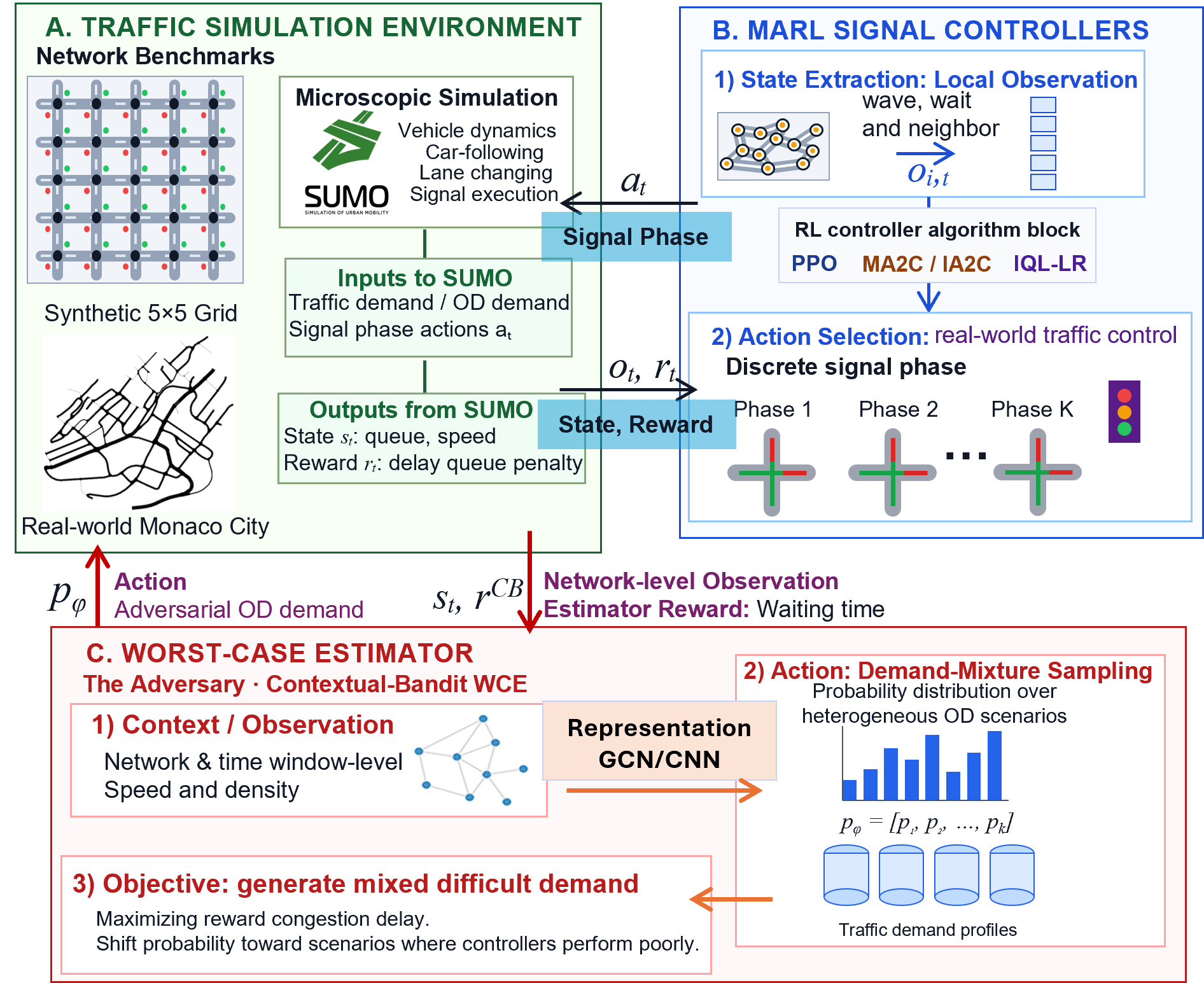}
    \caption{Schematic of the proposed DR-MARL training framework for intelligent intersection management. The diagram is adapted and extended from our previous work~\cite{pei2025distributionally} to reflect the broader set of traffic networks and learning algorithms considered in this paper.}
    \label{fig:overview}
\end{figure}

Reinforcement learning (RL) offers a compelling data-driven alternative, learning optimal policies directly through environmental interaction without explicit traffic models~\cite{sutton1998reinforcement}. While deep RL has substantially outperformed fixed-time control at isolated intersections~\cite{huang2023reinforcement}, scaling to network levels typically requires multi-agent reinforcement learning (MARL). In MARL, decentralized agents control individual intersections using local observations and neighborhood communications, demonstrating promising coordination on complex road networks~\cite{zhang2023learning}. However, a critical limitation persists: most existing MARL controllers are trained on nominal demand distributions and optimize solely for expected return~\cite{bukharin2023robust}. Consequently, they overfit to average-case scenarios and lack resilience against distributional shifts.

From an operational perspective, optimizing average performance alone is insufficient; traffic authorities prioritize travel time reliability under adverse conditions, such as extreme congestion or disrupted flows, where the predictability of the system is as critical as its efficiency \cite{jose2020travel}. Standard RL formulations lack explicit tail-performance guarantees, necessitating robust objectives~\cite{sagawa2020gdro}. While distributionally robust optimization (DRO) has succeeded in standard machine learning domains~\cite{sagawa2020gdro,hashimoto2018fairness}, its application to complex MARL traffic signal control remains unexplored.

Our recent work in~\cite{pei2025distributionally}, a precursor to this paper, introduced a distributionally robust multi-agent reinforcement learning (DR-MARL) framework for localized traffic grids by employing a contextual-bandit worst-case estimator (CB-WCE)~\cite{liu2025distributionallyrobustmultiagentreinforcement} that adaptively reweights traffic scenarios during training\footnote{The implementation and results for this DR-MARL framework are available at: \url{https://github.com/TravidP/DRMARL_CBWCE_Traffic_Controller}}. As shown in~\cref{fig:overview}, this estimator actively steers existing MARL controllers toward improved worst-case performance without altering their underlying policy architectures. However, that work was constrained to small-scale environments and limited algorithmic testing, leaving its scalability and general applicability an open question. In this present work, we extend our previous results in~\cite{pei2025distributionally} as follows.  Firstly, we do \emph{not} keep the estimator fixed during retraining (as shown later in Algorithm~\ref{alg:unified_cb_wce} in section~\ref{sec:method}): because the identity of the ``worst'' scenario can change as the controller improves, the CB-WCE continues to update so that it tracks the evolving worst-case conditions induced by the multi-agent policy. Secondly, this study significantly scales the methodology to complex environments, including an expanded $5\times5$ synthetic grid and the real-world Monaco Traffic network (MoST)~\cite{codeca2018monaco}. Finally, we demonstrate the framework's generality across multiple MARL paradigms: Independent Advantage Actor-Critic (IA2C)\cite{chu2019multi}, Multi-Agent Advantage Actor-Critic (MA2C) \cite{chu2019multi}, Proximal Policy Optimization (PPO) \cite{schulman2017proximal}, and Independent Q-Learning with Linear Regression (IQL-LR)\cite{tan1993multi}.

The main contributions of this work are summarized as follows:

\begin{enumerate}
    \item We implement a CB-WCE that co-evolves with a MARL traffic controller by adaptively reweighting demand scenarios during their simultaneous training, enabling distributionally robust learning under demand uncertainty.

    \item We demonstrate that the implemented CB-WCE framework is algorithm-agnostic and can be seamlessly integrated with existing RL traffic signal controllers across value-based, actor–critic, and policy-based methods, without requiring modifications to their architectures. Beyond synthetic $5 \times 5$ grids, we validate its scalability using the heterogeneous Monaco network, proving its efficacy in the real-world.
    \item Through extensive simulations, we show that retraining existing MARL controllers using the CB-WCE improves worst-case performance across demand scenarios while enhancing average-case performance. In particular, for the PPO architecture in the Monaco environment, robust retraining reduced the worst-case queue length by 74.39\% and improved the average-case network-wide queue length by 75.45\%. Notably, the framework exhibits robust zero-shot generalization, maintaining improvement under unseen demand distributions without requiring additional fine-tuning.

\end{enumerate}

The rest of the paper is organized as follows. Section~\ref{sec:related_work} surveys the current state of MARL for traffic systems and robustness techniques. Section~\ref{sec:problem_statement} provides the mathematical problem formulation and baseline descriptions, while Section~\ref{sec:method} provides a methodology of CB-WCE framework. We describe our experimental environment and training protocols in Section~\ref{sec:setup}. In Section~\ref{sec:results}, we evaluate the framework’s performance across synthetic and real-city benchmarks. Finally, we conclude in Section~\ref{sec:conclusion} with contributions and future research. 


\section{Related Work}
\label{sec:related_work}

RL has been extensively applied to traffic signal control, transitioning from isolated intersection management to large-scale, coordinated network operations. In this section, we review the evolution of MARL for traffic control and discuss recent advances in robust and DR-RL that motivate our proposed framework.

\subsection{MARL for Traffic Controller: Single and Multi-Agent}
\label{subsec:marl_tsc}
This section traces the progression of deep reinforcement learning for traffic signal control from isolated-intersection controllers to multi-agent coordination across road networks.

\subsubsection{Deep RL at Isolated Intersections}
Early applications of deep RL to traffic signal control focused on isolated intersections, leveraging deep neural networks to process high-dimensional state representations. The work in \cite{genders2016} introduced a Deep Q-Network (DQN) controller evaluated in high-fidelity simulation environments, using a Convolutional Neural Network (CNN)-based encoding of approaching traffic. Similarly, both value-based and policy-gradient methods using visual state representations are explored in \cite{mousavi2017}. Subsequently, an image-like state formulation, IntelliLight, is proposed in \cite{wei2018intellilight}, capturing lane occupancy, waiting time, and phase context to determine optimal phase switching. While highly effective locally, these single-agent formulations struggle with network-wide coordination and the non-stationarity induced by neighboring intersections.

\subsubsection{Multi-Agent Coordination}
To achieve scalable network-level control, the field shifted toward MARL, often leveraging actor-critic architectures to stabilize learning across multiple intersections. Initial efforts demonstrated that decentralized agents relying on local observations could implicitly coordinate across arterial networks \cite{van2016coordinated}. With the maturation of deep RL, explicit coordination mechanisms were introduced. In this context,  a decentralized DQN-based controller, PressLight, was proposed in \cite{wei2019presslight}, utilizing pressure-based traffic features for stability. Building on spatial dependencies, CoLight, proposed in \cite{wei2019colight}, utilized graph attention networks to model interactions at intersections, while traffic control using message passing was formulated in \cite{zhang2023learning}. Furthermore, a scalable deep MARL framework was introduced in \cite{chu2019multi} using local queue, waiting-time, and phase-timer observations alongside spatial discounting, setting a strong benchmark for grid networks.

Despite sophisticated coordination mechanisms, these MARL approaches typically optimize for expected returns under static, nominal demand distributions. Consequently, they are highly sensitive to traffic-demand variability and often suffer severe performance degradation under unseen or adversarial demand shifts \cite{9294623, huang2026robustefficientmultiagentreinforcement}. In addition, brute-force attempts to train across all possible traffic states demand impractical amounts of computational resources, hindering real-world deployment. This operational gap emphasizes the need for training paradigms that can efficiently enhance the resilience of MARL traffic controllers against distributional shifts in demand.
\subsection{Distributionally Robust RL in Traffic Controller}
\label{subsec:robust_marl_tsc}
This section reviews robustness-oriented reinforcement learning approaches for traffic signal control.
\subsubsection{Robust Reinforcement Learning under Uncertainty}
To address performance degradation under shifting environment dynamics, robustness in RL has traditionally been framed through robust Markov decision processes (MDPs) and risk-sensitive objectives \cite{Zhang2020}. In this context, the work of \cite{yamagata2024safe} provides an extensive survey of the safe and robust RL landscape. Early robust MDP formulations sought policies optimized against worst-case transition or reward uncertainties \cite{iyengar2005robust,nilim2005robust}, while risk-sensitive methods focused on variance-related criteria \cite{tamar2015cvar}. In continuous control and deep RL, domain randomization, adversarial perturbations \cite{pinto2017rarl}, and worst-trajectory sampling \cite{rajeswaran2017epopt} are frequently used to bias training toward difficult environments.

Recently, these concepts have begun to permeate traffic signal control. For instance, \cite{bukharin2023robust} proposed adversarial regularization to enforce policy Lipschitz continuity, stabilizing MARL traffic controllers against state perturbations. Similarly, the work in~\cite{bodagala2026uncertainty} incorporated belief-space optimization and explicit safety constraints to handle partial observability. However, these methods primarily focus on robustness to state perturbations or sensor noise rather than addressing macroeconomic distributional shifts in traffic demand, which fundamentally alter the underlying environment dynamics and require a different robustness paradigm.

\subsubsection{Distributionally Robust Optimization}
An alternative to adversarial noise injection is DRO, which adaptively controls the emphasis placed on different contexts or subpopulations during training \cite{kuhn2025distributionally}. Contextual-bandit formulations treat distinct environment configurations as inducing stochastic returns, allocating samples to optimize a robustness objective \cite{langford2008epochgreedy,dudik2011doubly,agarwal2014taming}. This aligns closely with Group DRO, which improves worst-group performance by dynamically upweighting underperforming scenarios \cite{sagawa2020gdro,hashimoto2018fairness}, while online algorithms for Distributionally Robust Markov Games (DRMGs) handle model uncertainties without prior offline data \cite{farhatsample}.

Extending this to MARL, recent work by \cite{liu2025distributionallyrobustmultiagentreinforcement} demonstrates that treating environmental variations as contexts and reweighting adverse instances during training can substantially elevate worst-case returns without compromising average performance in Amazon warehouse dispatching. \emph{However, to the best of the authors' knowledge, in the domain of traffic signal control, the explicit integration of dynamic traffic demand distributions to optimize robust RL policies remains unexplored.}

Motivated by these findings, our work bridges the gap between structured traffic demand and DR-RL. By treating distinct traffic demand regimes as contexts, we adopt a CB-WCE that adaptively samples difficult traffic conditions, seamlessly grafting distributional robustness onto established MARL traffic controllers.

\section{MARL Based Traffic Signal Control}
\label{sec:problem_statement}

This section introduces the traffic-signal control problem and the baseline learning algorithms evaluated in this work. We first formulate network-wide signal control as a MARL problem, specifying the environment dynamics, agent observations, action space with safety constraints, and reward structure. This problem formulation is shared across all methods considered. We model network-wide traffic signal control as a MARL problem on urban road networks simulated in \textsc{SUMO}~\cite{Krajzewicz2012SUMO}. It is instantiated on two distinct networks: a synthetic $5\times5$ grid and a real-world  subarea of Monaco City \cite{codeca2018monaco}. We model each signalized intersection as a single learning agent. Let $\mathcal{I} \coloneqq \{1,\dots,|\mathcal{I}|\}$ denote the set of controlled junctions, and let $t \in \{0,\dots,H-1\}$ index discrete \emph{decision steps}, at which signal actions are updated every $\Delta t$ seconds, within an episode of length $H$. 

\subsection{Action Space}
\label{subsubsec:actionspace_controller}

Traffic signal control is formulated using discrete, finite action spaces that specify which groups of vehicle movements are granted right-of-way at each decision step. At decision step $t$, the agent controlling intersection $i$ selects a signal phase $a_t^i$ from a predefined set of mutually exclusive and conflict-free phases. 
In the synthetic $5\times5$ grid network, all intersections share an identical geometry and phase design, resulting in a common action space $\mathcal{A}^i = \{0, 1, 2, 3, 4\}$ for all agents $i \in \mathcal{I}$. By contrast, in the Monaco city network, intersection geometries and permitted turning movements vary, and each intersection $i$ therefore operates on its own individual action space $\mathcal{A}^i = \{0, 1, \dots, |\mathcal{P}^i| - 1\}$, where $\mathcal{P}_i$ denotes the set of feasible signal phases defined by the local layout. Thus, at any time step $t$, the action taken by intersection $i$ is drawn from its specific action space, $a_t^i$.

In both environments, an action determines which vehicle movements receive right-of-way during the subsequent control interval, while all incompatible movements remain red.
Signal timing constraints and safety protocols are intrinsically managed by the simulation environment and remain exogenous to the learning process. In this framework, actions are executed over a fixed control interval of $\Delta t = 5$~s. While recent literature, such as \cite{aslani2017adaptive}, utilizes a longer interval of $\Delta t = 10$~s with a yellow phase of $t_y = 5$~s, this study adopts a reduced $\Delta t = 5$~s and a clearance sequence (comprising yellow and all-red intervals) of 2~s. This parameterization was selected to enhance the temporal resolution of the control policy, thereby increasing the system's adaptiveness to rapid fluctuations in traffic flow. 


\subsection{State Space}

To capture local traffic dynamics, agents rely on compact lane-level measurements. Following \cite{chu2019multi}, the primary observation includes two features: \emph{wave} (the number of approaching vehicles per incoming lane) and \emph{wait} (the cumulative delay of the lead vehicle). For real-world application, both metrics can be derived from data collected by on-road sensing units.

Formally, the base local observation for intersection $i$ at decision step $t$ is:
\begin{equation}
\bigl\{\text{wave}_t[l],\, \text{wait}_t[l]\bigr\}_{l \in \mathcal{L}_i},
\end{equation}
where $\mathcal{L}_i$ is the set of incoming lanes at intersection $i$. This full feature set is used in the synthetic $5\times5$ grid network. For the Monaco city network, to mitigate observation noise and improve learning stability under heterogeneous intersection geometries, the state is simplified to exclude the \emph{wait} feature.

To mitigate partial observability and non-stationarity in decentralized multi-agent settings, baseline controllers augment this local observation with neighbor information. Specifically, IQL-LR, PPO, and IA2C incorporate the \emph{wave} features of immediate neighbors. MA2C further integrates spatially discounted neighboring states and low-dimensional \emph{policy fingerprints} (recent action probabilities of adjacent agents). Table~\ref{tab:obs_summary} summarizes the observation components for each controller.


\begin{table}[t!]
\centering
\caption{Observation design for each controller and network.}
\label{tab:obs_summary}
\begin{tabularx}{\columnwidth}{@{}l c c Y@{}}
\toprule
\multirow{2}{*}{Controller} 
& \multicolumn{2}{c}{Local observation} 
& \multirow{2}{*}{Neighbor information} \\
\cmidrule(lr){2-3}
& 5 $\times$ 5 Grid & Monaco & \\
\midrule
IA2C   
& wave + wait 
& wave 
& wave \\

MA2C   
& wave + wait 
& wave 
& wave + policy \\

IQL-LR 
& wave + wait 
& wave 
& wave \\

PPO    
& wave + wait 
& wave 
& wave \\
\bottomrule
\end{tabularx}
\end{table}

Let $\mathbf{o}_t \coloneqq \{o_t^i\}_{i \in \mathcal{I}}$ denote the global joint traffic state and $\mathbf{a}_t \coloneqq \{a_t^i\}_{i \in \mathcal{I}}$, the joint action across all intersections. The environment follows a Markovian transition process:
\begin{equation}
\mathbf{o}_{t+1} \sim P_{\mathrm{env}}\bigl(\,\cdot \mid \mathbf{o}_t, \mathbf{a}_t \bigr),
\end{equation}
where the transition kernel $P_{\mathrm{env}}$ is governed by \textsc{SUMO}'s microscopic vehicle dynamics and the executed signal phases over the control interval $\Delta t$.

\subsection{Reward}



The reward function is formulated to be locally measurable and spatially decomposable, allowing for decentralized credit assignment. At each decision step $t$, every agent $i \in \mathcal{I}$ receives a post-decision signal $r_t^i$ that penalizes the congestion and delay accumulated on its incoming lanes during the control interval $\Delta t$. 

For the synthetic $5\times5$ grid, we define the reward as a weighted sum of spatial and temporal penalties:

\begin{flalign}
r_{t}^i = - \sum_{l \in \mathcal{L}_i} \big( \text{queue}_{t+\Delta t}[l] + a \cdot \text{wait}_{t+\Delta t}[l] \big) &&
\end{flalign}
Here, $\text{queue}_{t+\Delta t}[l]$ represents the lane's queue length in vehicles, and $\text{wait}_{t+\Delta t}[l]$ denotes the cumulative waiting time in seconds, both measured immediately following the control interval $\Delta t$. The intuition behind this dual-objective reward is to balance network-wide throughput with individual vehicle fairness: penalizing the queue length discourages spatial spillback, while the waiting time penalty prevents phase starvation where a small number of vehicles are delayed
indefinitely. The coefficient is set to $a=0.2$. This specific value is selected to balance the scale difference between the discrete vehicle count and the continuous temporal delay, effectively penalizing long-standing vehicles to prevent localized gridlock.

In the heterogeneous Monaco City network, the reward is simplified to strictly penalize queue length:
\begin{flalign}
r_{t}^i = - \sum_{l \in \mathcal{L}_i} \text{queue}_{t+\Delta t}[l]. &&
\end{flalign}
This simplification is motivated by the irregular topology of the real-world network. Incorporating waiting times in such asymmetric environments often introduces excessive reward variance due to the varying lengths and capacities of the links, which can destabilize multi-agent coordination. By isolating queueing as the primary penalty, the framework prioritizes the clearing of critical bottlenecks to maintain systemic stability.

In both environments, these post-decision rewards explicitly map the selected signal phase to its direct impact on subsequent traffic conditions.

\subsection{Baseline MARL Algorithms}
\label{subsec:baseline_algorithms}
Following the benchmark established by \cite{chu2019multi}, we use IA2C, MA2C, PPO, and IQL-LR as our baseline controllers \cite{chu2019multi,tan1993multi, schulman2017proximal}. To ensure diverse function approximation mechanisms, IA2C, MA2C, and PPO utilize deep neural networks, whereas IQL-LR employs linear regression. Because these four algorithms span value-based, policy-based, and actor--critic methods, they provide a comprehensive testbed to evaluate whether our distributionally robust framework can yield consistent improvements. Each algorithm uses a experience replay buffer to store observed transitions, and updates are performed using mini-batches sampled from this buffer. 

\section{Distributionally Robust MARL for Traffic Signal Control}
\label{sec:method}

Rather than optimizing for a single nominal pattern, we seek policies resilient across a finite set of representative scenarios. To achieve this, CB-WCE adaptively samples challenging demand mixtures during training, enhancing worst-case robustness, while preserving average performance. This section presents a DR training framework for network-level traffic signal control under heterogeneous demand.

\begin{algorithm}[t!]
\caption{Distributionally Robust MARL via Contextual-Bandit Worst-Case Estimator}
\label{alg:unified_cb_wce}

\KwIn{Demand profiles $\{\lambda^{(k)}\}_{k\in\mathcal{K}}$; 
Initial traffic policies $\{\pi_{\theta_0^i}\}_{i\in\mathcal{I}}$; 
Initial estimator $\mu_{\psi_0}$; 
Step size $\Delta t$; window length $T_{\mathrm{win}}$; 
Windows per episode $H_{\mathrm{W}}$; 
Learning rates $l_\theta, l_\psi$; 
Total episodes $N_{\mathrm{rollout}}$
}

\KwOut{Robust traffic policies $\{\pi_{\theta^{i\ast}}\}_{i\in\mathcal{I}}$; 
Optimized estimator $\mu_{\psi^\ast}$
}

\vspace{2mm}

Initialize estimator replay buffer $\mathcal{D}_\psi \gets \emptyset$\;
Initialize estimator parameters $\psi \gets \psi_0$\;

\For{intersection $i \in \mathcal{I}$}{
    Initialize controller replay buffer $\mathcal{D}_\theta^i \gets \emptyset$\;
    Initialize policy parameters $\theta^i \gets \theta_0^i$\;
}

\vspace{2mm}

\For{episode $k = 1, \dots, N_{\mathrm{rollout}}$}{

    Reset environment and initialize estimator state $s_0$\;

    \For{window $\tau = 1, \dots, H_{\mathrm{W}}$}{

        \tcp{Train Contextual-Bandit Worst-Case Estimator}
        Update estimator observation $s_\tau$\;
        Sample demand weights $w_\tau \sim \mu_\psi(\cdot \mid s_\tau)$\;
        Construct mixed demand $\lambda_\tau = \sum_{k\in\mathcal{K}} w_{\tau,k}\lambda^{(k)}$\;

        Initialize window reward $r_\tau^{\mathrm{CB}} \gets 0$\;

        \tcp{Train Traffic Controllers under Adversarial Demand}
\For{step $t = 1, \dots, T_{\mathrm{win}}$}{

    \For{intersection $i \in \mathcal{I}$}{
        Observe local state $o_t^i$\;
        Sample local action $a_t^i \sim \pi_{\theta^i}(\cdot \mid o_t^i)$\;
    }

    Construct joint action $A_t = \{a_t^i\}_{i\in\mathcal{I}}$\;
    Execute safe joint action $A_t$ for $\Delta t$\;

    \For{intersection $i \in \mathcal{I}$}{
        Obtain $r_t^i$ and next observation $o_{t+1}^i$\;

        $\mathcal{D}_\theta^i \gets \mathcal{D}_\theta^i 
        \cup \{(o_t^i, a_t^i, r_t^i, o_{t+1}^i)\}$\;
    }

    Update cumulative estimator reward $r_\tau^{\mathrm{CB}}$\;
}

        Observe next estimator state $s_{\tau+1}$\;
        $\mathcal{D}_\psi \gets \mathcal{D}_\psi \cup \{(s_\tau, w_\tau, r_\tau^{\mathrm{CB}}, s_{\tau+1})\}$\;

        \tcp{MARL Traffic signal Policy Update }
        \For{intersection $i \in \mathcal{I}$}{
            Sample mini-batch $B_\theta^i \sim \mathcal{D}_\theta^i$\;
            $\theta^i \gets \theta^i + l_\theta \nabla_{\theta^i} J_{\mathrm{MARL}}(\theta^i; B_\theta^i)$\;
        }
    }

    \tcp{Estimator Update}
    Sample mini-batch $B_\psi \sim \mathcal{D}_\psi$\;
    $\psi \gets \psi + l_\psi \nabla_\psi J_{\mathrm{CB}}(\psi; B_\psi)$\;
}
\Return $\{\pi_{\theta^{i}\ast} \gets \pi_{\theta^i}\}_{i\in\mathcal{I}}, \mu_{\psi^\ast} \gets \mu_\psi$\;

\end{algorithm}
\subsection{Distributionally Robust Performance Objective}
\label{subsec:robust-objective}

Traffic-signal control policies deployed in practice face substantial demand variability due to time-of-day effects, incidents, and routing shifts. Rather than optimizing under a single nominal demand distribution, we consider a finite set of representative scenarios \( \mathcal{K}=\{1,\dots,K\} \). Each scenario encodes a distinct traffic context (e.g., peak hours or spatial imbalances) obtained via synthetic generation, as detailed in Section~\ref{subsubsec:demand_groups}.

Let $\theta$ denote the parameter vector governing the multi-agent policy $\pi_\theta$. For a fixed policy parameterized by $\theta$ and a given traffic demand scenario $k \in \mathcal{K}$, the scenario-conditioned expected return is defined as:
\(
    J_k(\theta) = \mathbb{E}[R(\tau) \mid k, \pi_\theta]
    \label{eq:expected_return}
\),
where $\tau$ denotes a rollout trajectory and $R(\tau)$ is the cumulative reward (formulated as a negative proxy for delay). System performance across these scenarios is summarized by average and worst-case criteria:
\begin{equation}
J_{\mathrm{avg}}(\theta)
\;=\;
\frac{1}{K}\sum_{k\in\mathcal{K}} J_k(\theta),
\qquad
J_{\mathrm{worst}}(\theta)
\;=\;
\min_{k\in\mathcal{K}} J_k(\theta).
\label{eq:j-avg-worst}
\end{equation}

Standard MARL typically maximizes expected returns over a fixed sampling distribution. Consequently, controllers often overfit to frequently sampled regimes and degrade substantially under challenging or underrepresented scenarios \cite{huang2026robustefficientmultiagentreinforcement}. To mitigate this, we formulate a finite-support DR objective. Our goal is to maximize $J_{\mathrm{worst}}(\theta)$ to guard against the most adverse scenario, while preserving a strong $J_{\mathrm{avg}}(\theta)$ to avoid conservative, inefficient behavior. 

To achieve this balance, we introduce an adaptive training mechanism that dynamically shifts sampling emphasis toward scenarios where the current policy underperforms. The following section presents the CB-WCE used to produce these adaptive weights and couple them with MARL retraining.

\subsection{Contextual-Bandit Worst-Case Estimator}
\label{subsec:cb-estimator}

To improve the worst-case performance defined in~\eqref{eq:j-avg-worst}, we adopt a CB-WCE that operates on a slower timescale than the traffic-signal controllers. The CB-WCE acts as an adversary: given a fixed signal-control policy $\pi_\theta$, it selects traffic-demand patterns that \emph{maximize} network-wide congestion, thereby biasing subsequent updates of $\pi_\theta$ toward challenging conditions. Unlike the signal-control agents which actuate every step $\Delta t$ seconds, the CB-WCE updates at a window-level timescale, acting once per \emph{traffic window} of fixed duration $T_{\mathrm{win}}$.
\subsubsection{Observation}
\label{subsubsec:wce_observation}

 At the end of each window $\tau$, the recent network evolution is compressed into a fixed-format feature vector
\(
s_\tau \in \mathbb{R}^{2|\mathcal{I}|}.
\) 
For each intersection $i\in\mathcal{I}$, $s_\tau$ concatenates two aggregated scalars computed over the preceding window: the \emph{average speed} and the \emph{average density}.\footnote{Other congestion summaries could be used; we choose speed and density because they are readily available in microscopic simulation and provide complementary information about traffic conditions.} Stacking these per-intersection features provides a compact, scale-agnostic snapshot of emerging congestion.

To effectively extract spatial dependencies from this snapshot, the CB-WCE's neural network architecture is tailored to the underlying road topology. For the structurally homogeneous $5\times5$ grid, the feature vector is reshaped into a two-dimensional spatial matrix and processed using a CNN~\cite{lecun2002gradient} to capture localized congestion patterns. In contrast, for the highly irregular and heterogeneous topology of the Monaco city network, we represent the intersections as nodes in a graph and process the feature vector $s_\tau$ utilizing a Graph Convolutional Network (GCN)~\cite{kipf2016semi} to model complex, non-Euclidean spatial correlations.

\subsubsection{Action}
\label{subsubsec:wce_action}

Based on $s_\tau$, the CB-WCE outputs a probability distribution over the scenario set
\begin{flalign}
w_\tau \in \Delta^{K}
\;=\;
\left\{ w \in \mathbb{R}_{\ge 0}^{K} \;\middle|\; \sum_{k=1}^{K} w_k = 1 \right\} &&
\end{flalign}
with one component $w_{\tau,k}$ per traffic-demand scenario $k\in\mathcal{K}$. In our implementation, $w_\tau$ is produced by a neural network policy utilizing a softmax activation to enforce the simplex constraint. 

These selected weights parameterize the demand used during the \emph{next} traffic window of length $T_{\mathrm{win}}$ by forming a convex combination of scenario-specific demand descriptors. Denoting $\{\lambda^{(k)}\}_{k\in\mathcal{K}}$ as the base demand parameters for each scenario, the resulting mixed demand is
\begin{flalign}
\label{eq:lambda_mix}
\lambda_{\mathrm{mix}}(w_\tau) = \sum_{k\in\mathcal{K}} w_{\tau,k}\,\lambda^{(k)} &&
\end{flalign}
This mixed demand is instantiated in the simulator and held constant for the duration of the subsequent window, after which the CB-WCE receives a new observation and repeats the cycle.

\subsubsection{Reward}
\label{subsubsec:wce_reward_update}

During the ensuing window of length $T_{\mathrm{win}}$ under demand $\lambda_{\mathrm{mix}}(w_\tau)$, the traffic-signal controller executes its standard (faster) control loop. The CB-WCE subsequently receives a scalar congestion reward reflecting the severity of the induced traffic conditions. We use the network-wide cumulative waiting time:
\begin{flalign}
r^{\mathrm{CB}}_\tau
\;=\;
\sum_{t=1}^{T_{\mathrm{win}}}\; \sum_{v \in \mathcal{V}_t} \mathrm{wait\_time}_v(t) &&
\end{flalign}
where $\mathcal{V}_t$ is the set of vehicles present at time $t$, and $\mathrm{wait\_time}_v(t)$ denotes vehicle $v$'s waiting-time contribution at time $t$. The CB-WCE is trained to \emph{maximize} $r^{\mathrm{CB}}_\tau$ to synthesize ``hard'' scenarios, directly counteracting the traffic-signal controller's objective to minimize delay.

Parameterized by a stochastic policy $\mu_\psi(w \mid s)$, the CB-WCE is trained using window-level experience tuples $(s_\tau, w_\tau, r^{\mathrm{CB}}_\tau, s_{\tau+1})$ stored in a replay buffer $\mathcal{D}_\psi$, as shown in Algorithm~\ref{alg:unified_cb_wce}. While the controller parameters $\theta_0$ are fixed during this stage, $\psi$ is updated via a REINFORCE-style contextual-bandit objective. For a mini-batch $B$ sampled from $\mathcal{D}_\psi$, we maximize
\begin{equation}
J_{\mathrm{CB}}(\psi; B)
\;:=\;
\frac{1}{|B|}
\sum_{(s_\tau, w_\tau, r_\tau^{\mathrm{CB}}, s_{\tau+1}) \in B}
r_\tau^{\mathrm{CB}} \,\log \mu_\psi(w_\tau \mid s_\tau),
\end{equation}
and apply stochastic gradient ascent,
\[
\psi \;\gets\; \psi + l_\psi \nabla_\psi J_{\mathrm{CB}}(\psi; B).
\]
This yields a trained estimator policy $\mu_{\psi^\ast}$ that effectively prioritizes scenario mixtures inducing maximum delay.

\subsection{Distributionally Robust MARL via CB-WCE}
\label{subsec:drmarl}

We now describe how the CB-WCE is coupled to MARL training to obtain a DR controller. Starting from a baseline signal-control policy $\pi_{\theta_0}$ trained under the nominal demand setting (Section~\ref{subsubsec:demand_groups}), the CB-WCE provides an online mechanism to emphasize traffic-demand scenarios that are currently difficult for the controller. In contrast to our previous work~\cite{pei2025distributionally}, as shown in Algorithm~\ref{alg:unified_cb_wce}, we do \emph{not} keep the estimator fixed during retraining: because the identity of the ``worst'' scenario can change as the controller improves, the CB-WCE continues to update so that it tracks the evolving worst-case conditions induced by $\pi_\theta$.

Training proceeds on two interacting time scales. Each episode is divided into traffic windows of duration $T_{\mathrm{win}}$. At the start of window $\tau$, the CB-WCE observes the window-level state summary $s_\tau$ (Section~\ref{subsubsec:wce_observation}) and samples scenario weights
\(
w_\tau \sim \mu_{\psi}(\cdot \mid s_\tau),
\)
which define the mixed demand $\lambda_{\mathrm{mix}}(w_\tau)$ applied for the upcoming window. Within the window, the traffic-signal controller executes its standard control loop at interval $T_{\mathrm{ctrl}}$, collecting the usual per-step rewards and transitions. At the end of the window, we compute the CB-WCE reward $r^{\mathrm{CB}}_\tau$ (Section~\ref{subsubsec:wce_reward_update}) and the next CB-WCE observation $s_{\tau+1}$, and store the tuple $(s_\tau, w_\tau, r^{\mathrm{CB}}_\tau, s_{\tau+1})$ in $\mathcal{D}_\psi$. Over the full episode, the controller experience is stored in $\mathcal{D}_\theta$ and used to update $\theta$ with the same MARL objective as in the baseline (Section~\ref{subsec:baseline_algorithms}).

The key difference from standard training is that sampling is performed under an \emph{adaptive} mixture distribution: the CB-WCE is repeatedly updated (via the contextual-bandit objective in Section~\ref{subsubsec:wce_reward_update}) to assign higher probability to scenario mixtures that yield high delay under the \emph{current} controller $\pi_\theta$. Consequently, the controller is trained on a continually refreshed set of hard demand patterns, which biases learning toward improving worst-case performance while still allowing generalization across the scenario set.

\section{Experimental Setup}
\label{sec:setup}


This section describes the demand-generation procedures, training protocol, and evaluation methodology used to quantify the effect of worst-case demand training. We evaluate each trained controller under the same set of demand-group scenarios and report standard network-level performance metrics.\footnote{The source code and configuration files for the updated training protocol are available at: \url{https://github.com/TravidP/updated_worst_case_MARL}}


\subsection{Network}
\label{subsec:network}

 In the experiment, it is instantiated on two distinct networks: a synthetic $5\times5$ urban grid and a real-world traffic network derived from a subarea of Monaco City \cite{codeca2018monaco}.

\subsubsection{Synthetic $5\times 5$ Grid}
\begin{figure}[t]
    \centering
    \includegraphics[width=0.6\linewidth]{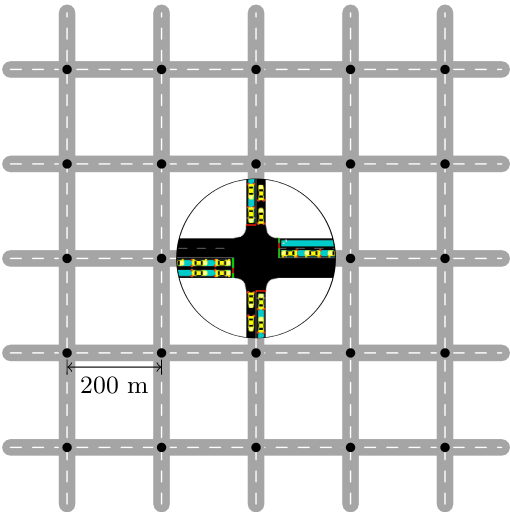}
    \caption{Synthetic $5 \times 5$ grid network, with an example intersection shown in the middle. Horizontal approaches have two lanes (one left-turn and one shared through/right) with a speed limit of $20$~m/s, while vertical approaches have a single mixed lane with a speed limit of $11$~m/s. The grid's regular structure allows for controlled experiments on coordination under homogeneous local control logic.}
    \label{fig:5x5_grid}
\end{figure}

The synthetic grid, shown in \cref{fig:5x5_grid}, comprises $25$ signalized intersections arranged in a regular lattice. 
This grid is a common benchmark for studying coordination under identical local control logic \cite{zhang2023learning,chu2019multi}.

\subsubsection{Monaco City Network.}
\begin{figure}[t]
    \centering
    \includegraphics[width=0.75\linewidth]{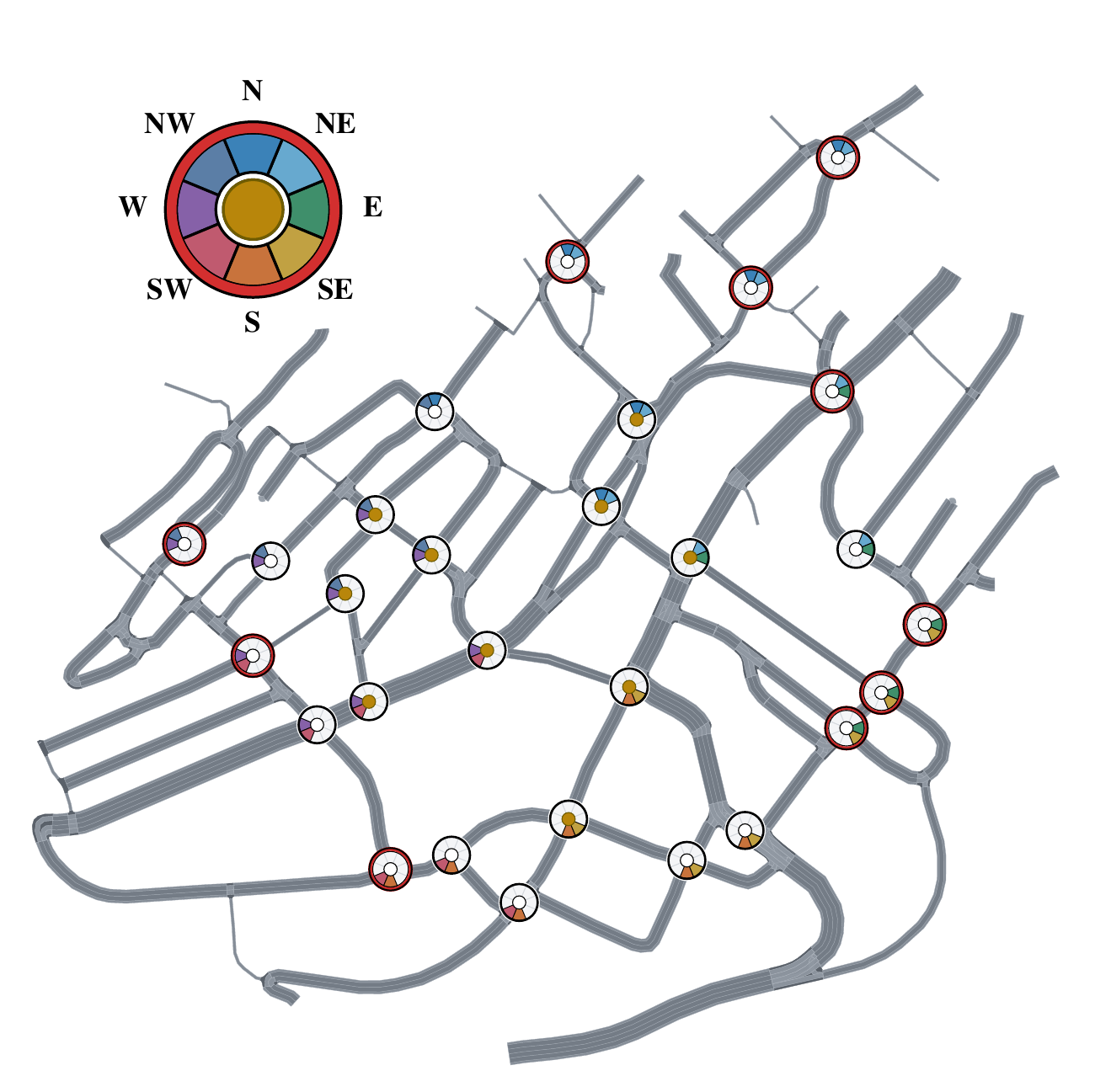}
    \caption{Monaco traffic network with controller locations overlaid by directional classification glyphs. Each glyph is a marker in which the filled outer sectors indicate the assigned directional classes (N, NE, E, SE, S, SW, W, NW), the filled inner circle denotes membership in the center group, and the red outer ring indicates the periphery class; multiple filled sectors indicate multi-directional assignment for the same controller.}
    \label{fig:monaco_grid}
\end{figure}

The Monaco benchmark, shown in \cref{fig:monaco_grid} is a real-world road network extracted from a central district of Monaco \cite{codeca2018monaco}. It comprises 30 signalized intersections connected by links with varying lengths, lane counts, speed limits, and turning configurations. To facilitate analysis, the map is partitioned into distinct directional groups based on the relative distances between locations. In contrast to the homogeneous grid, intersections in Monaco differ substantially in topology and feasible phase sets, resulting in heterogeneous local action spaces and non-uniform interaction patterns.

\subsection{Demand groups}
\label{subsubsec:demand_groups}

We define a finite set of $K=11$ synthetic demand groups, each specifying a distinct origin--destination (OD) inflow pattern on the network boundaries to represent qualitatively diverse traffic scenarios. In total, across the $K=11$ training demand groups, indexed as $k \in \{1, \dots, 11\}$, this set encompasses directional corridor flows (e.g., north--south and east--west), diagonal flows, center-focused loading (periphery-to-center and center-to-periphery), and a uniform demand distribution. We apply these scenarios to two distinct networks.

For the $5 \times 5$ grid network in \cref{fig:5x5_grid}, dominant directional flows are assigned an intensity of approximately $2000$~veh/h, yielding a total network demand of roughly $3000$~veh/h per scenario when combined with secondary background traffic. For the Monaco City benchmark (see \cref{fig:monaco_grid}), the network is partitioned into distinct spatial and directional groups based on relative distances. Following the dataset established in \cite{codeca2018monaco}, the total demand is maintained at a baseline of approximately $2383.3$~veh/h, but is systematically redistributed to reflect our variable OD patterns. During the simulation, vehicle departures are generated stochastically to preserve the macroscopic OD structure while ensuring that individual episodes differ in their microscopic vehicle realizations.

\subsubsection{Baseline training demand}
For baseline training, each episode cycles through all $K$ demand groups once. Within each window exactly one demand group is active, so that every episode exposes the controller to the full scenario family.

\subsubsection{Demand for robust retraining}
During robust retraining, the demand groups remain unchanged but the sequence of groups is no longer fixed. Instead, the CB-WCE selects mixture weights over the $K$ demand groups at the window time scale. Let $\mathbf{w}_\tau \in \Delta^{K-1}$ denote the weight vector produced by the CB-WCE at demand window $\tau$. The applied demand is then given by the convex combination, as shown in \cref{eq:lambda_mix}. By adaptively shifting probability mass toward demand groups that induce poor performance for the current controller, the CB-WCE biases training toward challenging traffic conditions while preserving the underlying scenario family.

\subsection{Training Protocol}
\label{subsec:training}
The framework evaluates $K=11$ distinct demand groups. Each group is active for a demand window of $T_{\mathrm{win}} = 600$~s (10 minutes). This duration is explicitly selected to emulate real-world traffic dynamics, operating under the practical assumption that a specific macroscopic traffic demand distribution typically persists for 10 minutes before undergoing a temporal shift. Consequently, the total duration of a single simulation episode is $H = 6600$~s. Within these windows, a fixed control interval of $\Delta t = 5$~s is maintained.

\subsubsection{Phase I: Baseline Training}
Baseline MARL controllers (IA2C, MA2C, IQL-LR, PPO) are trained for $10^6$ environment steps under a deterministic schedule. Each episode cycles sequentially through all $K=11$ predefined demand groups exactly once, yielding converged policies optimized for average-case returns.

\subsubsection{Phase II: CB-WCE Initialization}
With MARL parameters temporarily frozen, the CB-WCE is trained to exploit baseline vulnerabilities. Updating at the $T_{\mathrm{win}} = 600$~s timescale, the CB-WCE learns a policy $\mu_\psi(w \mid s)$ that outputs a probability distribution over the $K$ demand groups for 500 episodes. Each simulation episode is $H = 6600$~s, including $K=11$ predefined demand groups. By maximizing network congestion, it generates adversarial demand mixtures tailored to the controllers' weaknesses.

\subsubsection{Phase III: Robust Retraining}
MARL controllers are initialized with their Phase I parameters and co-evolved with the CB-WCE for 1,000 episodes. Instead of a fixed schedule, traffic demand is dynamically constructed every $T_{\mathrm{win}} = 600$~s using the adaptive mixtures sampled by the CB-WCE. Each simulation episode is $H = 6600$~s. The controllers update to minimize delay under these challenging conditions, while the CB-WCE continuously tracks their shifting vulnerabilities, driving the learning toward robust policies.

\subsection{Evaluation and Metrics}
\label{subsec:training_evaluation}

Trained controllers are evaluated across the $11$ training demand groups via $10$ independent rollouts of length $H_{\mathrm{eval}} = 3600$\,s, during which both the selected demand profile and the learned policy remain fixed. Furthermore, to independently test the controllers' zero-shot generalization to unseen traffic distributions, we include an additional demand pattern (designated as Group 12). For the $5\times 5$ grid, this pattern is derived from the Hangzhou traffic scenario \cite{wei2019survey,wei2019colight,zheng2019learning}, systematically mapped from its original $4\times4$ layout to the $5\times5$ benchmark. For the Monaco environment, it is the real life demand in the Monaco dataset \cite{codeca2018monaco}.

To assess performance, we compute two horizon-averaged, network-level metrics for each rollout: \emph{queue length}, defined as the total number of queued vehicles network-wide, and \emph{average speed}, representing the mean speed across all active vehicles. Finally, to quantify the efficacy of worst-case training, we directly compare the baseline controllers against their DR variants, with the resulting performance across all demand scenarios detailed in Section~\ref{sec:results}.

\begin{figure}[t!]
    \centering
    \includegraphics[width=0.5\textwidth]{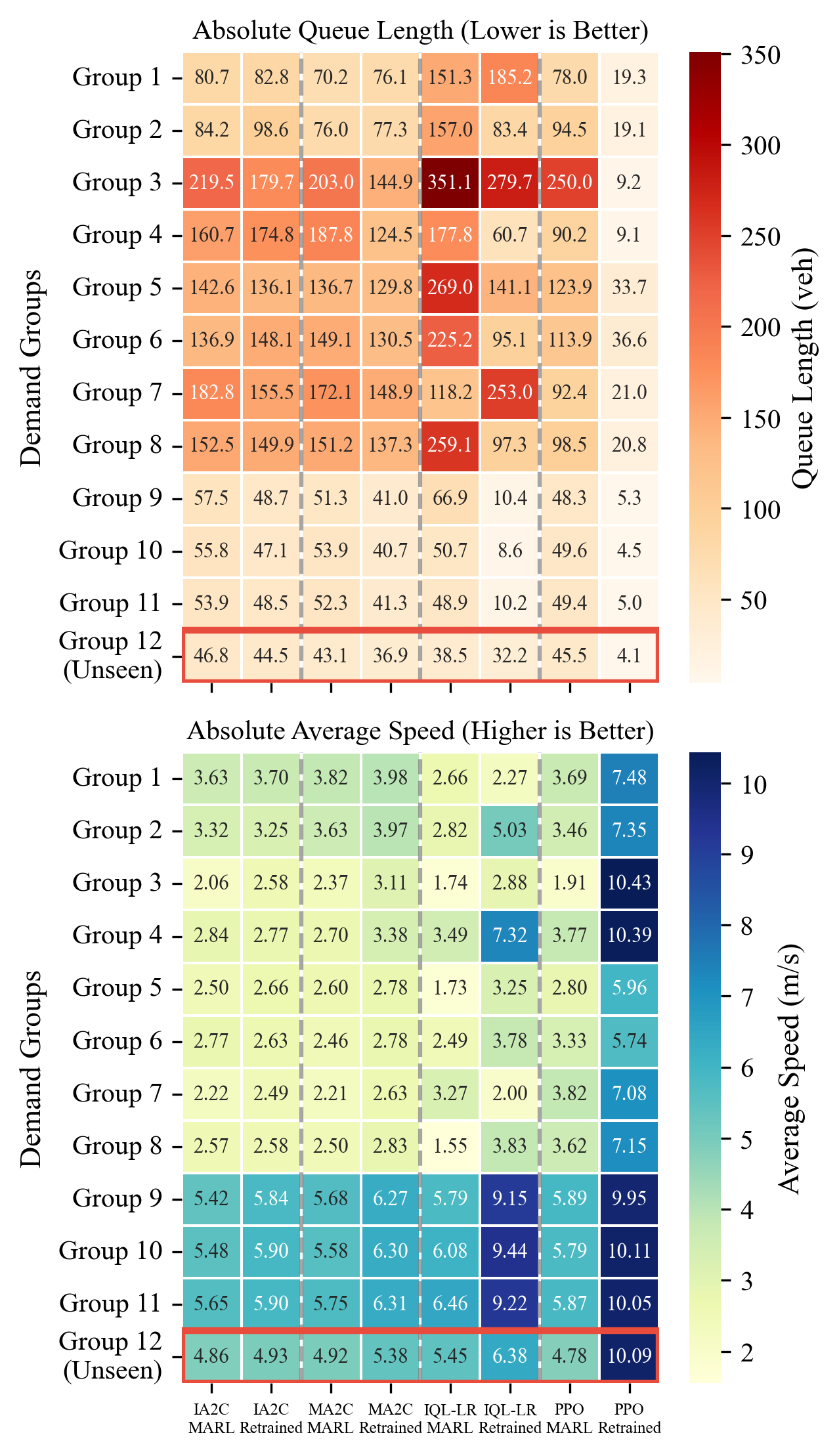} 

   \caption{Horizon- and rollout-averaged performance of baseline and distributionally robust (DR) controllers across demand groups in the $5 \times 5$ grid. The subplots depict the absolute average queue length in vehicles (Upper) and vehicle speed in m/s (Lower) per intersection. Queue length utilizes an orange-red color scale, where optimal performance (lower queue lengths) is represented by less red, while vehicle speed utilizes a yellow-green-blue scale where optimal performance (higher speeds) is indicated by dark blue. Four MARL architectures are evaluated in baseline and DR-retrained configurations, separated by dashed lines. Groups 1--11 represent training distributions, while Group 12 (red box) serves as an unseen environment for evaluating zero-shot generalization.}

    \label{Heatmap_Comparison_5x5}
\end{figure}

\begin{figure}[t!]
    \centering
    \includegraphics[width=0.44\textwidth]{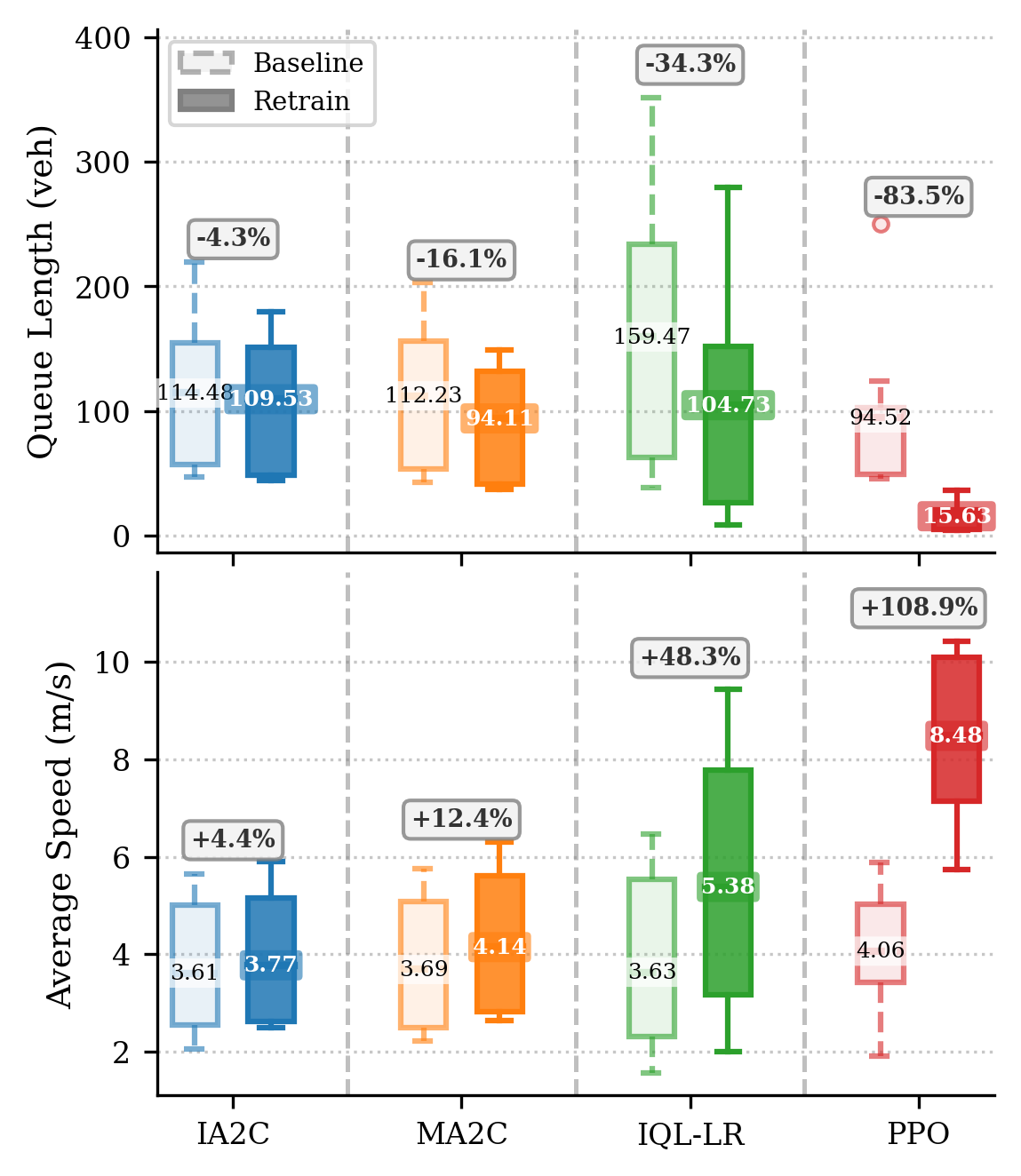} 
    \caption{Performance distributions of baseline (light fill, dashed) and DR-retrained (dark fill, solid) MARL controllers in the $5 \times 5$ grid. Boxplots evaluate Queue Length (Upper) and Average Speed (Lower) across four architectures. Numeric annotations denote mean values, and top grey boxes indicate the relative percentage change post-retraining. DR retraining consistently improves both metrics across all models, highlighted by an $83.46\%$ queue reduction and a $108.91\%$ speed increase for PPO.}

    \label{Box_Comparison_5x5}
\end{figure}

\begin{figure}[t!]
    \centering
    \includegraphics[width=0.47\textwidth]{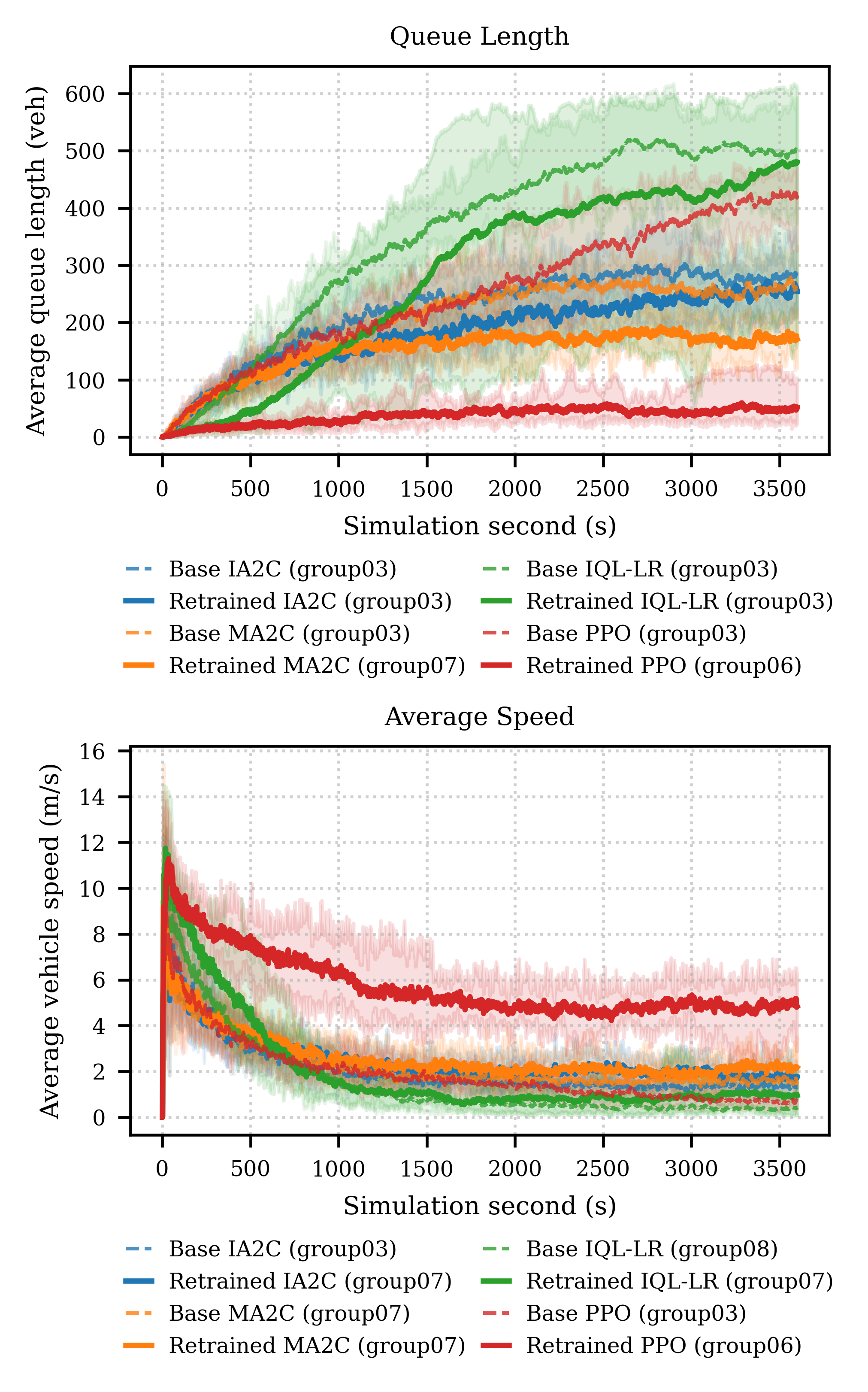} 
    \caption{Temporal evolution of worst-case queue length (upper) and average speed (lower) in the $5 \times 5$ grid over a \SI{3600}{\second} simulation horizon. Trajectories compare baseline MARL controllers (dashed) against their DR-retrained counterparts (solid), with shaded regions indicating min-max variance across evaluation rollouts. Retraining effectively flattens performance variance and yields a global performance lift, frequently shifting the worst-case demand distribution (e.g., from Group 03 to Group 06 for PPO).}

    \label{Worst_case_5x5}
\end{figure}
\section{Results and Discussion}
\label{sec:results}

This section presents the empirical evaluation of the proposed DR training framework for multi-agent traffic signal control. We first report the results on the synthetic $5\times5$ grid network, followed by experiments on the Monaco traffic network. Finally, we discuss the observed performance trends and their implications for worst-case robustness, average performance, and the general applicability of the proposed approach. For every controller and demand group, we compute the horizon-averaged queue length and average speed, and report the mean of these metrics across the $10$ rollouts.

\subsection{Results: $5\times5$ grid network}
\label{subsec:results}

The horizon-averaged performance across demand groups is summarized in \cref{Heatmap_Comparison_5x5,Box_Comparison_5x5}. These plots compare the baseline MARL controllers with their DR counterparts. Across all evaluated algorithms, baseline performance varies significantly depending on the demand group. However, robust retraining consistently yields improvements in both average capabilities and worst-case bottleneck scenarios (depicted in \cref{Worst_case_5x5}). The following subsections detail the specific numerical improvements for each algorithm.

\subsubsection{IA2C}
Across all groups, the overall mean queue length decreased by $4.32\%$ (from $114.98$ to $109.03$), while mean speed improved by $4.38\%$ (from $3.53$ to $3.68\,\mathrm{m/s}$). Regarding worst-case performance, the maximum queue length (observed in group 3 for both variants) dropped by $18.13\%$ to $179.72$ vehicles. The worst-case speed also saw a $20.87\%$ relative improvement, rising from $2.06\,\mathrm{m/s}$ (group 3) to $2.49\,\mathrm{m/s}$ (shifting to group 7).

\subsubsection{MA2C}
Overall, after retraining, this represents a $16.15\%$ reduction in average queue length (from $112.90$ to $94.94$) and a $12.42\%$ increase in average speed (from $3.52$ to $4.15\,\mathrm{m/s}$). The worst-case queue length shifted from group 3 ($203.02$) to group 7 ($148.93$), marking a substantial $26.64\%$ reduction. The worst-case speed, consistently found in group 7, improved by $19.00\%$ from $2.21$ to $2.63\,\mathrm{m/s}$.

\subsubsection{IQL-LR}
On average, queue lengths fell by $34.33\%$ (from $159.64$ to $104.15$) and speeds surged by $48.22\%$ (from $3.63$ to $5.21\,\mathrm{m/s}$). The worst-case queue length remained in group 3 but dropped by $20.33\%$ to $279.70$ vehicles. The worst-case speed improved by $29.03\%$, increasing from $1.55\,\mathrm{m/s}$ (group 7) to $2.00\,\mathrm{m/s}$ (group 7).

\subsubsection{PPO}
The robust retraining corresponds to an $83.46\%$ drop in average queue length (from $94.51$ to $16.71$) and a $108.91\%$ increase in average speed (from $3.86$ to $8.65\,\mathrm{m/s}$). The worst-case queue length plunged by $86.53\%$, shifting from group 3 ($250.00$) to group 5 ($33.68$). Similarly, the worst-case speed improved by an exceptional $200.52\%$, rising from $1.91\,\mathrm{m/s}$ (group 3) to $5.74\,\mathrm{m/s}$ (group 6).

\begin{figure}[t!]
    \centering
    \includegraphics[width=0.5\textwidth]{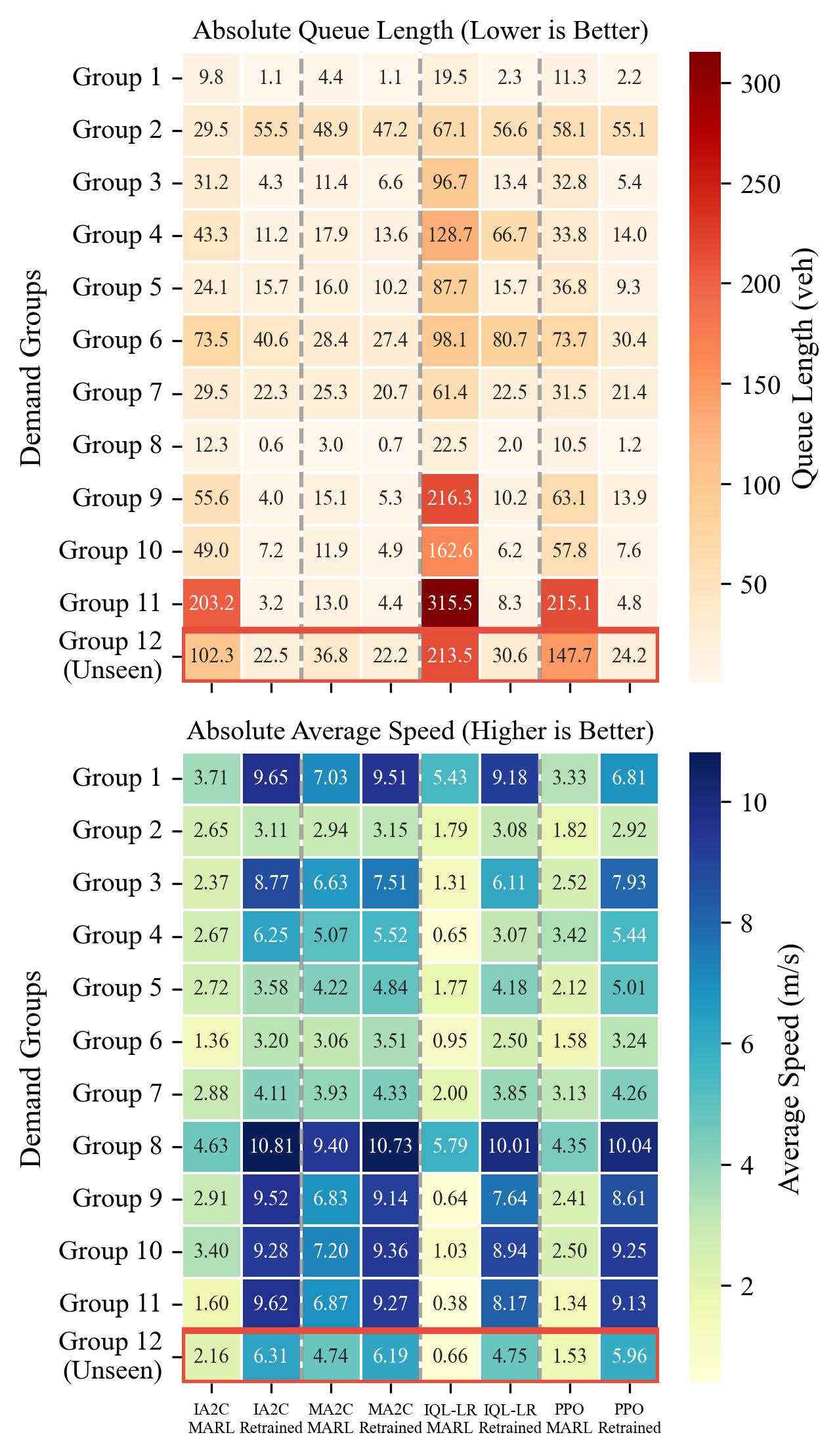} 
    \caption{Horizon- and rollout-averaged performance of baseline and distributionally robust (DR) controllers across demand groups in the Monaco city network. The subplots depict the absolute average queue length in vehicles (upper) and vehicle speed in m/s (lower) per intersection. Queue length utilizes an orange-red color scale where optimal performance (lower queue lengths) is represented by less red, while vehicle speed utilizes a yellow-green-blue scale where optimal performance (higher speeds) is indicated by dark blue. Four MARL architectures are evaluated in baseline and DR-retrained configurations, separated by dashed lines. Groups 1--11 represent training distributions, while Group 12 (red box) serves as an unseen environment for evaluating zero-shot generalization.}

    \label{Heatmap_Comparison_Monaco}
\end{figure}

\begin{figure}[t!]
    \centering
    \includegraphics[width=0.44\textwidth]{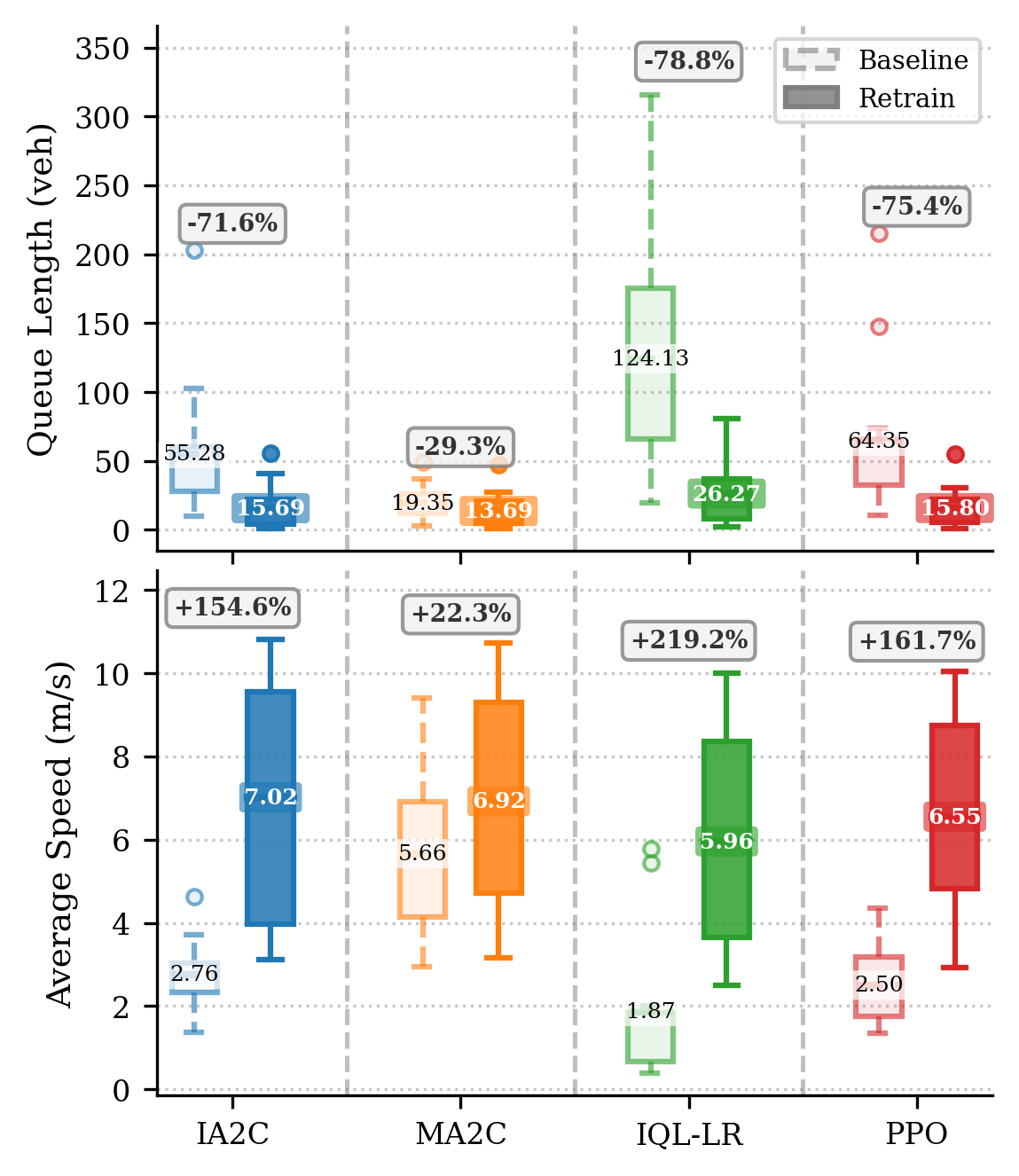} 
    \caption{Performance distributions of baseline (light fill, dashed) and DR-retrained (dark fill, solid) MARL controllers in the Monaco city network. Boxplots evaluate Queue Length (upper) and Average Speed (lower) across four architectures. Numeric annotations denote mean values, and top grey boxes indicate the relative percentage change post-retraining. DR retraining consistently improves both metrics across all models, highlighted by a $78.84\%$ queue reduction and a $219.16\%$ speed increase for IQL-LR.}

    \label{Box_Comparison_Monaco}
\end{figure}

\begin{figure}[t!]
    \centering
    \includegraphics[width=0.47\textwidth]{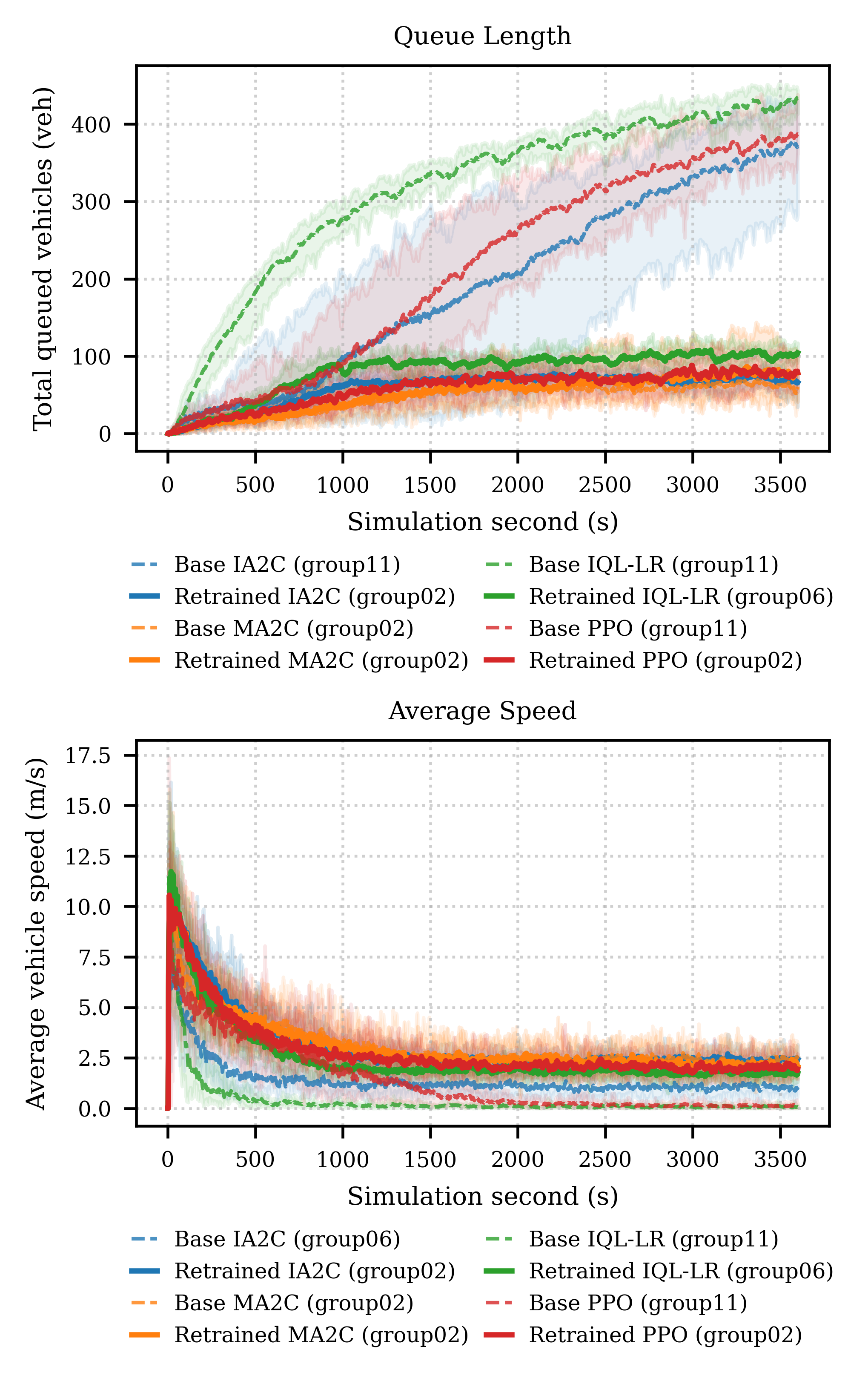} 
    \caption{Temporal evolution of worst-case queue length (upper) and average speed (lower) in the Monaco city network over a \SI{3600}{\second} simulation horizon. Trajectories compare baseline MARL controllers (dashed) against their DR-retrained counterparts (solid), with shaded regions indicating min-max variance across evaluation rollouts. Retraining effectively flattens performance variance and yields a global performance lift, frequently shifting the worst-case demand distribution (e.g., from Group 11 to Group 02 for PPO).}
    \label{Worst_case_Monaco}
\end{figure}

\subsection{Results: Monaco city network}\label{subsec:results_monaco}

To assess the scalability and effectiveness of the proposed framework under realistic, irregular network structures, we evaluate the controllers on the Monaco City benchmark. This real-world road network introduces significant heterogeneity in intersection layouts, lane counts, and turning structures, which poses a greater challenge for multi-agent coordination compared to the synthetic grid. Similar to the previous study, we compare the baseline MARL controllers with their DR counterparts across 11 demand groups and additional unseen group 12, as summarized in \cref{Heatmap_Comparison_Monaco,Box_Comparison_Monaco,Worst_case_Monaco}. The real-time traffic congestion after optimization is illustrated in \cref{fig:congestion_comparison}. The specific numerical enhancements for each algorithm are detailed below.

\subsubsection{IA2C}
Overall, the mean queue length decreased by $71.62\%$ (from $55.28$ to $15.69$), while the average speed improved by $154.55\%$ (from $2.76$ to $7.02\,\mathrm{m/s}$). The worst-case queue length shifted from group 11 ($203.24$) to group 2 ($55.49$), representing a $72.70\%$ reduction. The worst-case speed improved by $128.68\%$, rising from $1.36\,\mathrm{m/s}$ (group 6) to $3.11\,\mathrm{m/s}$ (group 2).

\subsubsection{MA2C}
The retraining corresponds to a $29.25\%$ reduction in average queue length (from $19.35$ to $13.69$) and a $22.29\%$ increase in average speed (from $5.66$ to $6.92\,\mathrm{m/s}$). The maximum queue length consistently occurred in group 2, reducing by $3.47\%$ to $47.20$ vehicles. The worst-case speed, also in group 2, improved by $7.14\%$ to $3.15\,\mathrm{m/s}$.

\subsubsection{IQL-LR}
Overall, average queue lengths dropped by $78.84\%$ (from $124.13$ to $26.27$), and speeds surged by $219.16\%$ (from $1.87$ to $5.96\,\mathrm{m/s}$). The worst-case queue shifted from group 11 ($315.46$) to group 6 ($80.72$), a $74.41\%$ reduction. Similarly, worst-case speed improved by $557.89\%$, rising from $0.38\,\mathrm{m/s}$ (group 11) to $2.50\,\mathrm{m/s}$ (group 6). 

\subsubsection{PPO}
\begin{figure}[t!]
    \centering
    \begin{tikzpicture}
        \node[anchor=south west, inner sep=0pt] (map) at (0,0) {\includegraphics[width=0.75\linewidth]{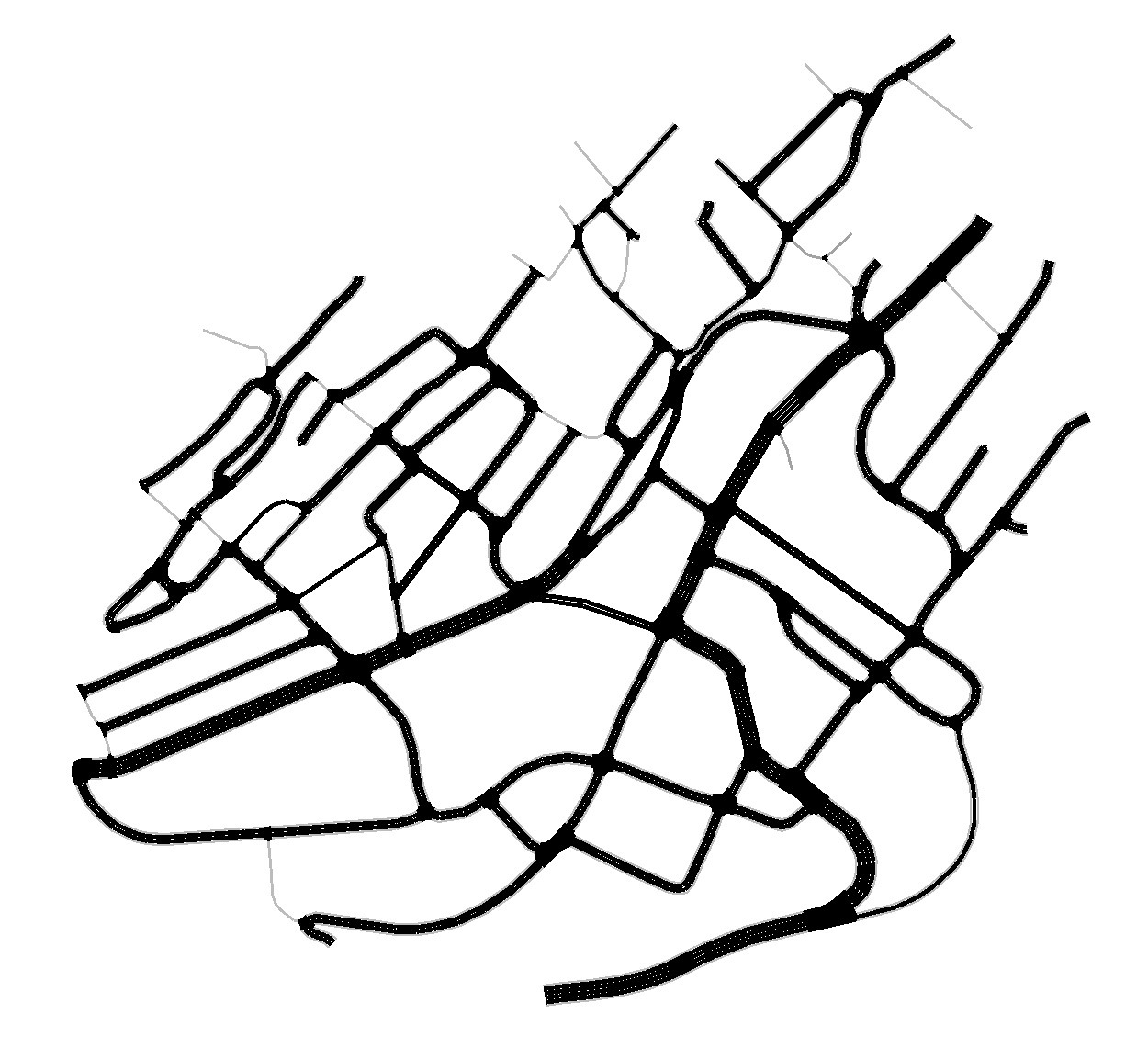}};

        \begin{scope}[x={(map.south east)},y={(map.north west)}]
        \node[circle, draw=red, very thick, minimum size=0.1\linewidth, inner sep=0pt] 
                (zoomnode) at (0.57, 0.48) {};
            
            \node[inner sep=0pt, anchor=north west, draw=black, thick] (baseline) at (0.02, 0.98) {
                \includegraphics[width=0.32\linewidth]{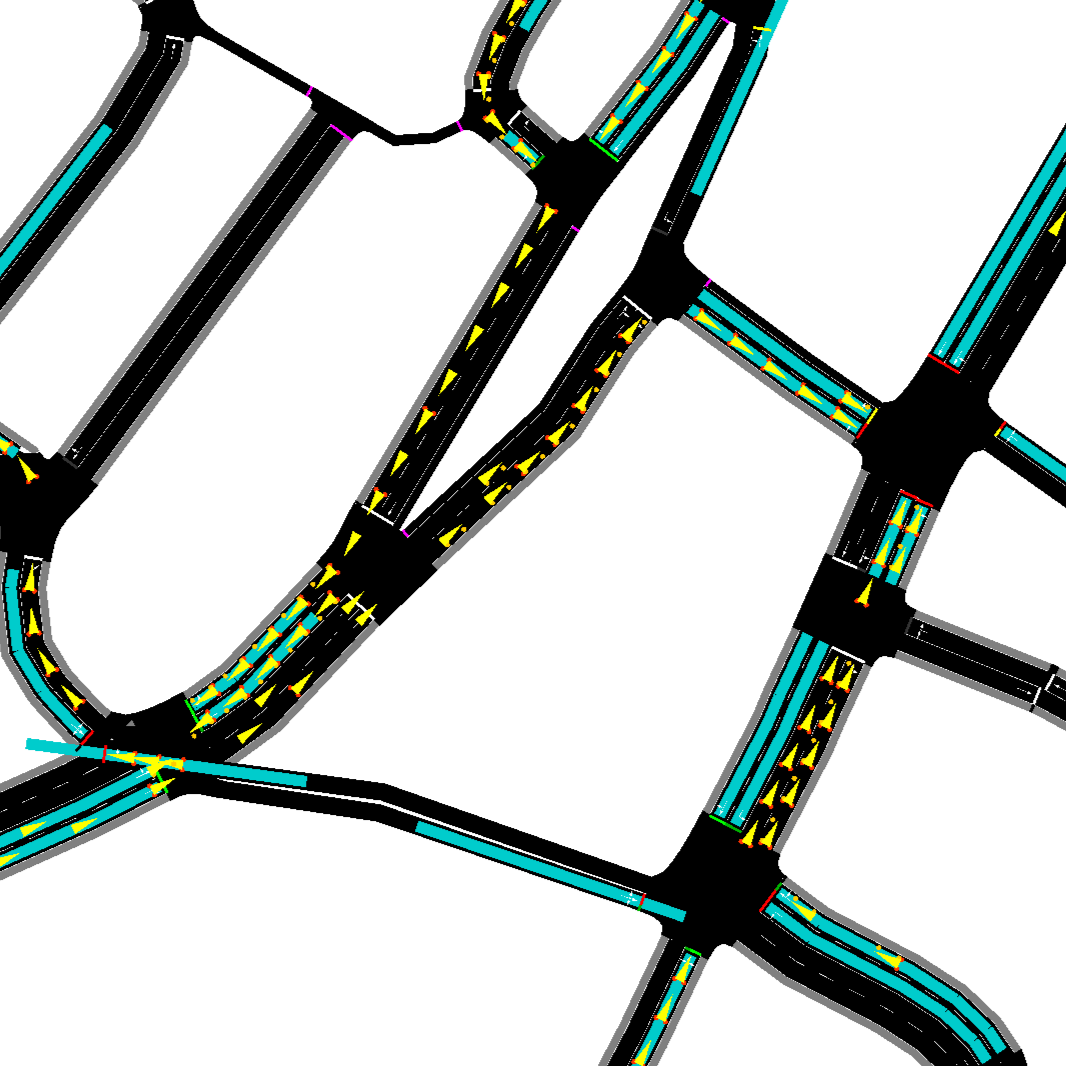}
            };
            \node[fill=white, draw=black, inner sep=2pt, anchor=north west, font=\scriptsize] 
                at (baseline.north west) {Baseline (Congested)};

            \node[inner sep=0pt, anchor=north west, draw=black, thick] (retrained) at (0.02, 0.48) {
                \includegraphics[width=0.32\linewidth]{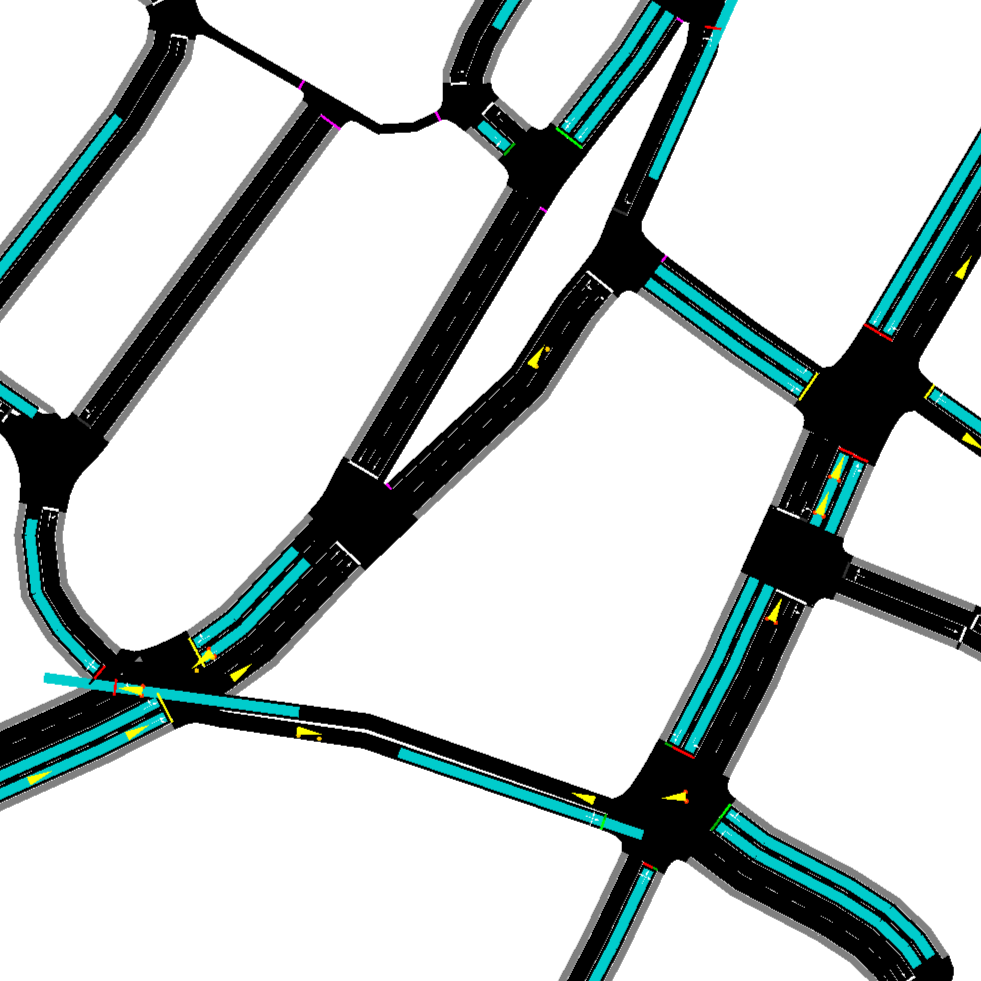}
            };
            \node[fill=white, draw=black, inner sep=2pt, anchor=north west, font=\scriptsize] 
                at (retrained.north west) {Retrained (Smooth)};

            \draw[-stealth, red, thick, dashed] (zoomnode) -- (baseline.east);
            \draw[-stealth, red, thick, dashed] (zoomnode) -- (retrained.east);

        \end{scope}
    \end{tikzpicture}
    \caption{Performance comparison at the Monaco city intersection using the PPO algorithm: Baseline MARL vs. Retrained MARL. This specific location was selected for evaluation as it represents a critical bottleneck characterized by peak congestion levels.}
    \label{fig:congestion_comparison}
\end{figure}
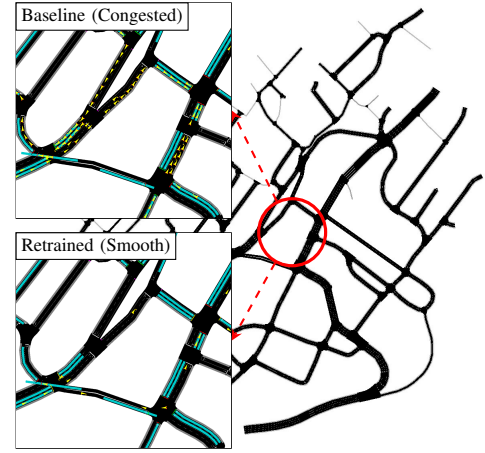

After robust retraining, it brings a $75.45\%$ decrease in overall average queue length (from $64.35$ to $15.80$) and a $161.74\%$ boost in average speed (from $2.50$ to $6.55\,\mathrm{m/s}$). The worst-case queue length shifted from group 11 ($215.10$) to group 2 ($55.08$), dropping by $74.39\%$. The worst-case speed improved by $141.79\%$, increasing from $1.34\,\mathrm{m/s}$ (group 11) to $3.24\,\mathrm{m/s}$ (group 2). \cref{fig:congestion_comparison} illustrates the reduction in congestion within the unseen group after retraining PPO using a worst-case estimator.

As observed in the other scenarios, the exceptional relative increases in speed (exceeding $100\%$) are largely a statistical artifact of the baseline controller's degraded performance under specific traffic conditions. The baseline model struggles to learn a generalized policy capable of accommodating diverse traffic scenarios simultaneously. As a result, the initial speeds for these challenging groups are near zero, meaning that standard absolute improvements in traffic flow translate into inflated relative percentage gains.

\subsection{Discussion}
\label{subsec:discussion}

The empirical evaluation across both the $5\times5$ grid and the Monaco City network reveals critical vulnerabilities in standard average-case training for multi-agent traffic signal control, while demonstrating the profound efficacy of the proposed DR framework. By analyzing the performance disparities between baseline and robustly retrained controllers, several key insights emerge.

\subsubsection{Topological Sensitivity and Scalability}
Our analysis reveals that different network topologies struggle with varying traffic distributions due to their unique structural characteristics. For instance, the $5\times5$ topology experiences significant performance degradation when handling the group 3 traffic distribution. This distribution features a high demand from the west to the east (2 lanes), which creates a bottleneck at the central intersection of the grid. Furthermore, the Monaco City topology struggles significantly with group 11. When demand shifts, this spatial-temporal overfitting causes catastrophic failures, such as the baseline IQL-LR degrading to $0.38\,\mathrm{m/s}$ in Group 11. The root cause of this lies in the topology's inherent structural asymmetry. Under a high-load uniform distribution (Group 11), the few critical bridge links connecting different network regions quickly become bottlenecks. The DR framework prevents this collapse, confirming that complex geography amplifies the penalty of demand uncertainty. Consequently, distributionally robust optimization is not merely a performance enhancement, but a prerequisite for ensuring stability and scalability when deploying these controllers across larger, increasingly heterogeneous urban networks.


\subsubsection{Observation Space and Generality}
The results highlight a critical relationship between observation space, network topology, and performance. During initial baseline evaluations in the Monaco network (\cref{Box_Comparison_Monaco}), MA2C exhibits superior performance and stability, an advantage driven by its explicit integration of neighbor policies and spatial states. However, following robust retraining, algorithms utilizing smaller, localized state representations achieve comparable performance levels. This indicates that with effective robust training, extensive neighbor observations are not strictly necessary to reach a high performance ceiling. This dynamic is further elucidated by the synthetic $5\times 5$ grid results (\cref{Box_Comparison_5x5}). Although MA2C initially outperforms IA2C and IQL-LR, it underperforms compared to PPO even in the initial baseline. Furthermore, despite robust retraining, MA2C fails to match the performance ceilings of both IQL-LR and PPO. This limitation stems from the $5\times 5$ grid's dense and uniform topology, which generates highly redundant neighbor observations. Such spatial data overload creates a learning bottleneck for MA2C, hindering further improvement despite robust retraining. In contrast, Monaco's irregular topology naturally restricts the available neighbor information, mitigating this redundancy and allowing MA2C to achieve a post-retraining performance on par with the other baseline algorithms.

These contrasting results demonstrate that merely expanding the observation space is not a substitute for optimal state design. Overloading the state space with dense neighbor data can inadvertently constrain the policy's upper bounds. Instead, observation features must be carefully formulated, allowing the DR framework to teach the policy how to isolate the most critical, localized signals that prevent systemic failure.

Crucially, despite these fundamental differences in observation design, the CB-WCE consistently improved performance across actor--critic, value-based, and policy-gradient architectures. This confirms that the framework is robust and algorithm-agnostic. Furthermore, CB-WCE accelerates training procedure. By actively prioritizing worst-case demand groups, the framework focuses precisely on critical state-action spaces, avoiding the redundant simulation of easily managed episodes.

\subsubsection{Global Performance and Zero-Shot Generalization}
The proposed framework improves controller efficacy across virtually \textit{all} demand distributions by training on the worst demand distributions. Rather than compromising average performance to hedge against extremes, training against adversarial demand mixtures yielded substantial system-wide enhancements. For example, the overall average queue length of all algorithms dropped between $29.25\%$ and $78.84\%$ in the Monaco network.
As the framework aggressively optimizes against primary vulnerabilities, the identity of the worst-case demand group shifts, as shown in \cref{Worst_case_Monaco} (e.g., IQL-LR's worst-case queue shifting from Group 10 to Group 5 in Monaco). This dynamic is a natural and expected outcome of the training methodology: by continuously targeting and optimizing for the current worst-performing scenario, the policy improves its handling of that specific condition until a different scenario takes its place at the bottom of the performance distribution. Because these newly established worst-case metrics substantially outperform the original baseline vulnerabilities, the framework effectively flattens performance variance across the uncertainty set while delivering an uncompromised, global performance lift.

Crucially, this system-wide enhancement shows potential to extend beyond the training distribution, suggesting a possibility for zero-shot generalization. In our evaluation, group 12 served as a completely unseen test dataset withheld during training. Despite having no prior exposure to these specific spatial-temporal distributions, as shown in \cref{Heatmap_Comparison_Monaco,Heatmap_Comparison_5x5}, all algorithmic architectures exhibited improved performance on group 12 post-retraining. Specifically for the Monaco scenario (\cref{fig:congestion_comparison}), the retrained MARL model significantly improves unseen group traffic performance, exhibiting fewer queued vehicles at identical intersections after the same simulation duration. Instead, by optimizing for worst-case bottlenecks, the controllers appear to learn more fundamental traffic-clearing dynamics. However, while these results demonstrate a promising possibility for zero-shot generalization against novel demand profiles, confirming the full extent and reliability of this capability requires further investigation.

\subsubsection{Practical Implications for Real-World Deployment}

From a practical traffic engineering perspective, the framework's value is validated by its performance in worst-case and unseen scenarios. Time-series analysis (\cref{Worst_case_5x5,Worst_case_Monaco}) reveals that while baseline models suffer from unbounded queue growth leading to gridlock, our DR-retrained models maintain stability. Crucially, this robustness translates to unseen traffic patterns (\cref{fig:congestion_comparison}), where the model effectively mitigates congestion, confirming its potential for real-world application.

Ultimately, the practical utility of this framework lies in its direct adaptability to empirical traffic data. While discrete demand groups serve as proxies in this study, traffic management agencies can construct custom uncertainty sets using historical sensor arrays. By feeding city-specific temporal dynamics using data-driven macroscopic clustering. (e.g., rush hour peaks, weekend events, adverse weather anomalies) into the DR framework, one can autonomously identify and optimize against localized network vulnerabilities, preemptively immunizing urban infrastructure against congestion collapse.

\section{Conclusion and Future Work}
\label{sec:conclusion}

This paper proposed a DR training framework utilizing CB-WCE for multi-agent traffic signal control. Empirical evaluations across a synthetic grid and the real-world Monaco City network demonstrate that baseline MARL is highly vulnerable to spatial-temporal demand shifts. By proactively targeting adversarial demand mixtures, our framework prevents unbounded queue growth and systemic gridlock. The CB-WCE approach consistently enhances both worst-case robustness and average-case efficiency across diverse architectures (IA2C, MA2C, IQL-LR, PPO), while showing strong potential for zero-shot generalization. 

Future work will first benchmark the DR framework against established real-world adaptive controllers and evaluate the performance bounds and computational efficiency of extended baseline training. Second, we aim to investigate alternative state-action representations to address partial observability in sensor-degraded environments, while extending the framework to multi-objective control for co-optimizing throughput, travel times, and vehicular emissions. Third, policy resilience must be evaluated under severe spatiotemporal disruptions, including traffic accidents and unannounced road closures. Finally, integrating dynamic macroscopic traffic assignment is critical to account for induced demand. Because optimized signal control improves local capacity, it inevitably alters global user behavior, attracting greedy routing and shifting socio-economic activity patterns. Modeling these coupled macroscopic-microscopic dynamics is essential to prevent congestion rebound and ensure long-term, sustainable network user equilibrium.
\balance

{\scriptsize
\bibliographystyle{IEEEtran}
\bibliography{references}

\begin{thebibliography}{10}
\providecommand{\url}[1]{#1}
\csname url@samestyle\endcsname
\providecommand{\newblock}{\relax}
\providecommand{\bibinfo}[2]{#2}
\providecommand{\BIBentrySTDinterwordspacing}{\spaceskip=0pt\relax}
\providecommand{\BIBentryALTinterwordstretchfactor}{4}
\providecommand{\BIBentryALTinterwordspacing}{\spaceskip=\fontdimen2\font plus
\BIBentryALTinterwordstretchfactor\fontdimen3\font minus \fontdimen4\font\relax}
\providecommand{\BIBforeignlanguage}[2]{{%
\expandafter\ifx\csname l@#1\endcsname\relax
\typeout{** WARNING: IEEEtran.bst: No hyphenation pattern has been}%
\typeout{** loaded for the language `#1'. Using the pattern for}%
\typeout{** the default language instead.}%
\else
\language=\csname l@#1\endcsname
\fi
#2}}
\providecommand{\BIBdecl}{\relax}
\BIBdecl

\bibitem{barth2008real}
M.~Barth and K.~Boriboonsomsin, ``Real-world carbon dioxide impacts of traffic congestion,'' \emph{Transportation Research Record}, vol. 2058, no.~1, pp. 163--171, 2008.

\bibitem{stanaway2018global}
J.~D. Stanaway, A.~Afshin \emph{et~al.}, ``Global, regional, and national comparative risk assessment of 84 behavioural, environmental and occupational, and metabolic risks or clusters of risks for 195 countries and territories, 1990--2017: A systematic analysis for the global burden of disease study 2017,'' \emph{The Lancet}, vol. 392, no. 10159, pp. 1923--1994, 2018.

\bibitem{dijkstra2021applying}
L.~Dijkstra, A.~J. Florczyk, S.~Freire, T.~Kemper, M.~Melchiorri, M.~Pesaresi, and M.~Schiavina, ``Applying the degree of urbanisation to the globe: A new harmonised definition reveals a different picture of global urbanisation,'' \emph{Journal of Urban Economics}, vol. 125, p. 103312, 2021.

\bibitem{Hunt1982SCOOT}
P.~Hunt, D.~Robertson, R.~Bretherton, and M.~C. Royle, ``The {SCOOT} on-line traffic signal optimisation technique,'' \emph{Traffic Engineering \& Control}, vol.~23, no.~4, pp. 190--192, 1982.

\bibitem{Luk1984SCATSCOOT}
J.~Luk, ``Two traffic-responsive area traffic control methods: {SCAT} and {SCOOT},'' \emph{Traffic Engineering \& Control}, vol.~25, no.~1, pp. 14--22, 1984.

\bibitem{gartner1982demand}
N.~H. Gartner, \emph{Demand-Responsive Decentralized Urban Traffic Control}.\hskip 1em plus 0.5em minus 0.4em\relax US Department of Transportation, Research and Special Programs Administration, 1982.

\bibitem{henry1984prodyn}
J.-J. Henry, J.~L. Farges, and J.~Tuffal, ``The {PRODYN} real time traffic algorithm,'' in \emph{Control in Transportation Systems}.\hskip 1em plus 0.5em minus 0.4em\relax Elsevier, 1984, pp. 305--310.

\bibitem{stevanovic2010atcs}
A.~Stevanovic, ``Adaptive traffic control systems: Domestic and foreign state of practice,'' National Cooperative Highway Research Program (NCHRP), Synthesis 403, 2010.

\bibitem{pei2025distributionally}
S.~Pei, J.~Borger, A.~Kosay, M.~O. Sayin, and S.~Ahmed, ``Distributionally robust multi-agent reinforcement learning for intelligent traffic control,'' \emph{IFAC-PapersOnLine}, 2026, accepted to the 23rd IFAC World Congress, Busan, Republic of Korea. arXiv:2512.18558.

\bibitem{sutton1998reinforcement}
R.~S. Sutton and A.~G. Barto, \emph{Reinforcement Learning: An Introduction}.\hskip 1em plus 0.5em minus 0.4em\relax Cambridge, MA: MIT Press, 1998.

\bibitem{huang2023reinforcement}
S.-C. Huang, K.-E. Lin, C.-Y. Kuo, L.-H. Lin, M.~O. Sayin, and C.-W. Lin, ``Reinforcement-learning-based job-shop scheduling for intelligent intersection management,'' in \emph{Proceedings of the Design, Automation \& Test in Europe Conference \& Exhibition (DATE)}, 2023, pp. 1--6.

\bibitem{zhang2023learning}
Y.~Zhang, Z.~Yu, J.~Zhang, L.~Wang, T.~H. Luan, B.~Guo, and C.~Yuen, ``Learning decentralized traffic signal controllers with multi-agent graph reinforcement learning,'' \emph{IEEE Transactions on Mobile Computing}, vol.~23, no.~6, pp. 7180--7195, 2023.

\bibitem{bukharin2023robust}
A.~Bukharin, Y.~Li, Y.~Yu, Q.~Zhang, Z.~Chen, S.~Zuo, C.~Zhang, S.~Zhang, and T.~Zhao, ``Robust multi-agent reinforcement learning via adversarial regularization: Theoretical foundation and stable algorithms,'' in \emph{Advances in Neural Information Processing Systems (NeurIPS)}, vol.~36, 2023, pp. 68\,121--68\,133.

\bibitem{jose2020travel}
A.~Jose and S.~Ram, ``Travel time reliability to airport: Review and assessment,'' \emph{Transportation Research Procedia}, vol.~48, pp. 2771--2783, 2020.

\bibitem{sagawa2020gdro}
S.~Sagawa, P.~W. Koh, T.~Hashimoto, and P.~Liang, ``Distributionally robust neural networks for group shifts: On the importance of regularization for worst-case generalization,'' in \emph{Proceedings of the 8th International Conference on Learning Representations (ICLR)}, 2020.

\bibitem{hashimoto2018fairness}
T.~Hashimoto, M.~Srivastava, H.~Namkoong, and P.~Liang, ``Fairness without demographics in repeated loss minimization,'' in \emph{Proceedings of the 35th International Conference on Machine Learning (ICML)}, 2018, pp. 1929--1938.

\bibitem{liu2025distributionallyrobustmultiagentreinforcement}
G.~Liu, S.~Iloglu, M.~Caldara, J.~W. Durham, and M.~M. Zavlanos, ``Distributionally robust multi-agent reinforcement learning for dynamic chute mapping,'' in \emph{Proceedings of the 42nd International Conference on Machine Learning (ICML)}, 2025, pp. 38\,722--38\,743.

\bibitem{codeca2018monaco}
L.~Codeca and J.~H{\"a}rri, ``{Monaco SUMO Traffic (MoST)} scenario: A {3D} mobility scenario for cooperative {ITS},'' \emph{EPiC Series in Engineering}, vol.~2, pp. 43--55, 2018.

\bibitem{chu2019multi}
T.~Chu, J.~Wang, L.~Codec{\`a}, and Z.~Li, ``Multi-agent deep reinforcement learning for large-scale traffic signal control,'' \emph{IEEE Transactions on Intelligent Transportation Systems}, vol.~21, no.~3, pp. 1086--1095, 2019.

\bibitem{schulman2017proximal}
J.~Schulman, F.~Wolski, P.~Dhariwal, A.~Radford, and O.~Klimov, ``Proximal policy optimization algorithms,'' \emph{arXiv:1707.06347}, 2017.

\bibitem{tan1993multi}
M.~Tan \emph{et~al.}, ``Multi-agent reinforcement learning: Independent vs. cooperative agents,'' in \emph{Proceedings of the 10th International Conference on Machine Learning (ICML)}, 1993, pp. 330--337.

\bibitem{genders2016}
W.~Genders and S.~Razavi, ``Using a deep reinforcement learning agent for traffic signal control,'' \emph{arXiv:1611.01142}, 2016.

\bibitem{mousavi2017}
S.~S. Mousavi, M.~Schukat, and E.~Howley, ``Traffic light control using deep policy-gradient and value-function-based reinforcement learning,'' \emph{IET Intelligent Transport Systems}, vol.~11, no.~7, pp. 417--423, 2017.

\bibitem{wei2018intellilight}
H.~Wei, G.~Zheng, H.~Yao, and Z.~Li, ``{IntelliLight}: A reinforcement learning approach for intelligent traffic light control,'' in \emph{Proceedings of the 24th ACM SIGKDD International Conference on Knowledge Discovery \& Data Mining (KDD)}, 2018, pp. 2496--2505.

\bibitem{van2016coordinated}
E.~Van~der Pol and F.~A. Oliehoek, ``Coordinated deep reinforcement learners for traffic light control,'' in \emph{Proceedings of the NIPS Workshop on Learning, Inference and Control of Multi-Agent Systems}, vol.~8, 2016, pp. 21--38.

\bibitem{wei2019presslight}
H.~Wei, C.~Chen, G.~Zheng, K.~Wu, V.~Gayah, K.~Xu, and Z.~Li, ``Presslight: Learning max pressure control to coordinate traffic signals in arterial network,'' in \emph{Proceedings of the 25th ACM SIGKDD International Conference on Knowledge Discovery \& Data Mining (KDD)}, 2019, pp. 1290--1298.

\bibitem{wei2019colight}
H.~Wei, N.~Xu, H.~Zhang, G.~Zheng, X.~Zang, C.~Chen, W.~Zhang, Y.~Zhu, K.~Xu, and Z.~Li, ``Colight: Learning network-level cooperation for traffic signal control,'' in \emph{Proceedings of the 28th ACM International Conference on Information and Knowledge Management (CIKM)}, 2019, pp. 1913--1922.

\bibitem{9294623}
C.~Wu, Z.~Ma, and I.~Kim, ``Multi-agent reinforcement learning for traffic signal control: Algorithms and robustness analysis,'' in \emph{Proceedings of the 23rd IEEE International Conference on Intelligent Transportation Systems (ITSC)}, 2020, pp. 1--7.

\bibitem{huang2026robustefficientmultiagentreinforcement}
S.-Y. Huang, H.-C. Chang, Y.-C. Chen, T.-H. Wei, I.-H. Yeh, S.-Y. Kuan, C.-Y. Wang, H.-H. Lee, and I.-C. Wu, ``A robust and efficient multi-agent reinforcement learning framework for traffic signal control,'' \emph{arXiv:2603.12096}, 2026.

\bibitem{Zhang2020}
K.~Zhang, T.~Sun, Y.~Tao, S.~Genc, S.~Mallya, and T.~Basar, ``Robust multi-agent reinforcement learning with model uncertainty,'' \emph{arXiv:2010.05445}, 2020.

\bibitem{yamagata2024safe}
T.~Yamagata and R.~Santos-Rodriguez, ``Safe and robust reinforcement learning: Principles and practice,'' \emph{arXiv:2403.18539}, 2024.

\bibitem{iyengar2005robust}
G.~Iyengar, ``Robust dynamic programming,'' \emph{Mathematics of Operations Research}, vol.~30, no.~2, pp. 257--280, 2005.

\bibitem{nilim2005robust}
A.~Nilim and L.~El~Ghaoui, ``Robust control of markov decision processes with uncertain transition matrices,'' \emph{Operations Research}, vol.~53, no.~5, pp. 780--798, 2005.

\bibitem{tamar2015cvar}
A.~Tamar, Y.~Glassner, and S.~Mannor, ``Policy gradients with variance related risk criteria,'' in \emph{Advances in Neural Information Processing Systems (NeurIPS)}, 2015, pp. 1653--1661.

\bibitem{pinto2017rarl}
L.~Pinto, J.~Davidson, R.~Sukthankar, and A.~Gupta, ``Robust adversarial reinforcement learning,'' in \emph{Proceedings of the 34th International Conference on Machine Learning (ICML)}, 2017.

\bibitem{rajeswaran2017epopt}
A.~Rajeswaran, K.~Lowrey, E.~Todorov, and S.~Kakade, ``{EPOpt}: Learning robust neural network policies using model ensembles,'' in \emph{Proceedings of the 5th International Conference on Learning Representations (ICLR)}, 2017.

\bibitem{bodagala2026uncertainty}
J.~Bodagala and B.~Bodagala, ``Uncertainty-aware counterfactual traffic signal control with predictive safety and starvation-avoidance constraints using vision-based sensing,'' \emph{arXiv:2602.07784}, 2026.

\bibitem{kuhn2025distributionally}
D.~Kuhn, S.~Shafiee, and W.~Wiesemann, ``Distributionally robust optimization,'' \emph{Acta Numerica}, vol.~34, pp. 579--804, 2025.

\bibitem{langford2008epochgreedy}
J.~Langford and T.~Zhang, ``Epoch-greedy algorithm for contextual bandits,'' in \emph{Advances in Neural Information Processing Systems (NeurIPS)}, 2008.

\bibitem{dudik2011doubly}
M.~Dud{\'\i}k, J.~Langford, and L.~Li, ``Doubly robust policy evaluation and learning,'' in \emph{Proceedings of the 28th International Conference on Machine Learning (ICML)}, 2011.

\bibitem{agarwal2014taming}
A.~Agarwal, D.~Hsu, S.~Kale, J.~Langford, L.~Li, and R.~E. Schapire, ``Taming the monster: A fast and simple algorithm for contextual bandits,'' in \emph{Proceedings of the 31st International Conference on Machine Learning (ICML)}, 2014.

\bibitem{farhatsample}
Z.~U. Farhat, D.~Ghosh, G.~K. Atia, and Y.~Wang, ``Sample-efficient distributionally robust multi-agent reinforcement learning via online interaction,'' in \emph{Proceedings of the 14th International Conference on Learning Representations (ICLR)}, 2026.

\bibitem{Krajzewicz2012SUMO}
D.~Krajzewicz, J.~Erdmann, M.~Behrisch, and L.~Bieker, ``Recent development and applications of {SUMO} -- simulation of urban {MObility},'' \emph{International Journal on Advances in Systems and Measurements}, vol.~5, no. 3 \& 4, pp. 128--138, 2012.

\bibitem{aslani2017adaptive}
M.~Aslani, M.~S. Mesgari, and M.~Wiering, ``Adaptive traffic signal control with actor-critic methods in a real-world traffic network with different traffic disruption events,'' \emph{Transportation Research Part C: Emerging Technologies}, vol.~85, pp. 732--752, 2017.

\bibitem{lecun2002gradient}
Y.~LeCun, L.~Bottou, Y.~Bengio, and P.~Haffner, ``Gradient-based learning applied to document recognition,'' \emph{Proceedings of the IEEE}, vol.~86, no.~11, pp. 2278--2324, 2002.

\bibitem{kipf2016semi}
T.~N. Kipf and M.~Welling, ``Semi-supervised classification with graph convolutional networks,'' \emph{arXiv:1609.02907}, 2016.

\bibitem{wei2019survey}
H.~Wei, G.~Zheng, V.~Gayah, and Z.~Li, ``A survey on traffic signal control methods,'' \emph{arXiv:1904.08117}, 2019.

\bibitem{zheng2019learning}
G.~Zheng, Y.~Xiong, X.~Zang, J.~Feng, H.~Wei, H.~Zhang, Y.~Li, K.~Xu, and Z.~Li, ``Learning phase competition for traffic signal control,'' in \emph{Proceedings of the 28th ACM International Conference on Information and Knowledge Management (CIKM)}, 2019, pp. 1963--1972.

\end{thebibliography}
}

\begin{IEEEbiography}[{\includegraphics[width=1in,height=1.25in,clip,keepaspectratio]{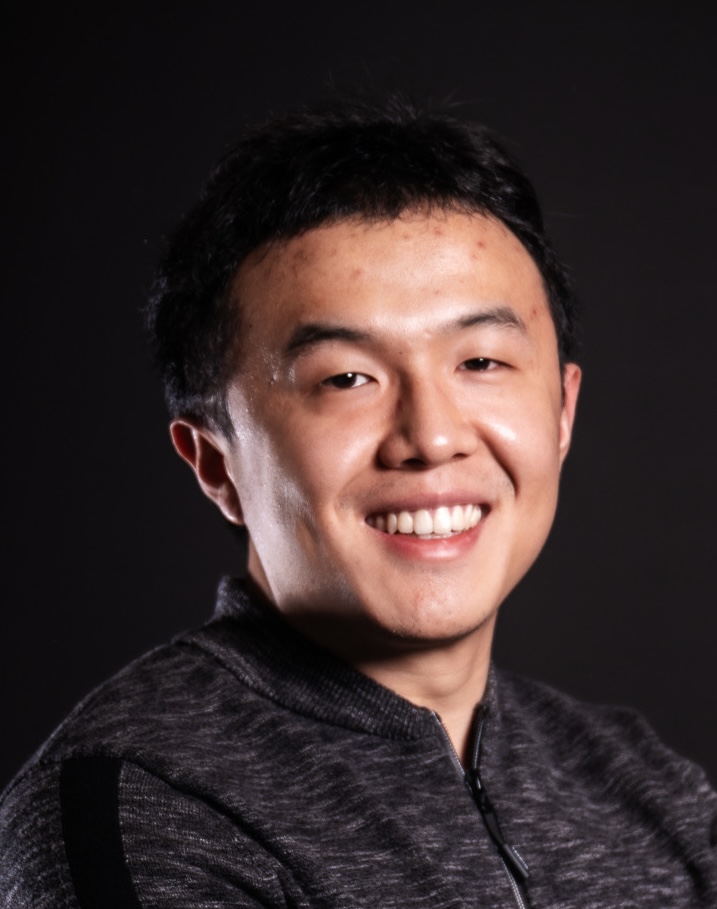}}]{Shuwei Pei} is currently pursuing the Ph.D. degree with the Faculty of Science and Engineering at the University of Groningen, Groningen, The Netherlands. Previously, he received the B.S. and M.S. degrees in Vehicle Engineering from the University of Science and Technology Beijing, Beijing, China, in 2021 and 2024, respectively. 
His research interests include intelligent transportation systems, multi-agent reinforcement learning and cooperative autonomous driving. His current work focuses on developing coordinated control strategies for urban traffic signal management, multi-lane traffic stabilization, and high-level decision-making for autonomous vehicle fleets.
\end{IEEEbiography}

\begin{IEEEbiography}[{\includegraphics[width=1in,height=1.25in,clip,keepaspectratio]{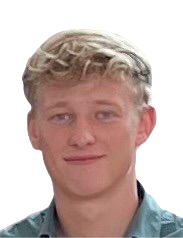}}]{Joran Borger} is currently pursuing a Master’s degree in Mechanical Engineering at the University of Groningen, Groningen, The Netherlands. He previously obtained his Bachelor’s degree in Mechanical Engineering from Hanze University of Applied Sciences, Groningen, The Netherlands, in 2023.
During his Master’s studies, he conducted research on the application of reinforcement learning in traffic signal control. He is currently completing his graduation research on the analysis of helical wire thread insertion processes, where machine learning techniques are applied to torque--angle data to investigate the quality and behavior of helical wire thread insertions.
\end{IEEEbiography}

\begin{IEEEbiography}[{\includegraphics[width=1in,height=1.25in,clip,keepaspectratio]{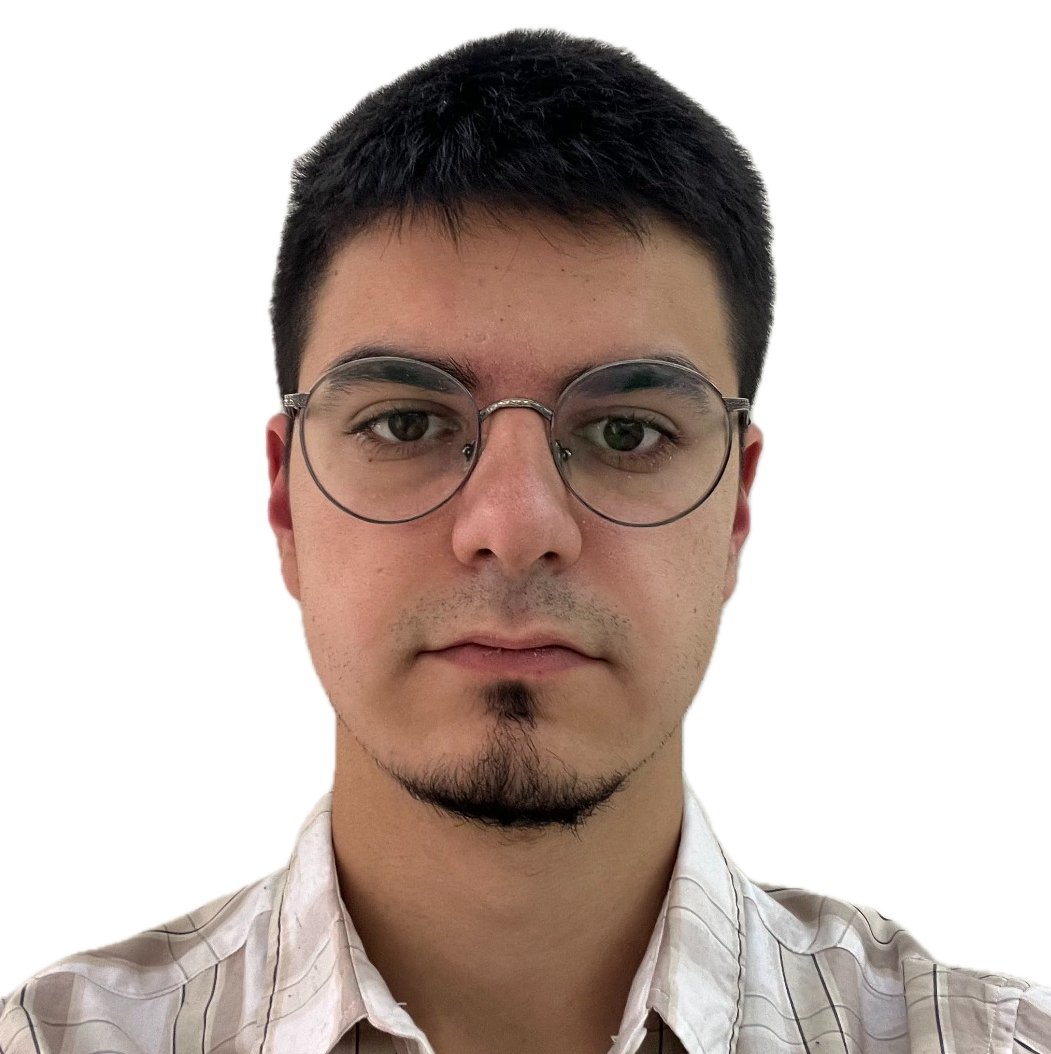}}]{Arda Kosay} is currently pursuing a Master's degree in Electrical and Electronics Engineering at Bilkent University, Ankara, Türkiye. He received his B.S. degree in the same department from Bilkent University, Ankara, Türkiye, in 2025. His research interests include developing learning-based and game-theoretic foundations for coordination, autonomy, and decision-making in dynamic multi-agent systems, with applications to intelligent transportation systems and networked control.
\end{IEEEbiography}

\begin{IEEEbiography}[{\includegraphics[width=1in,height=1.25in,clip,keepaspectratio]{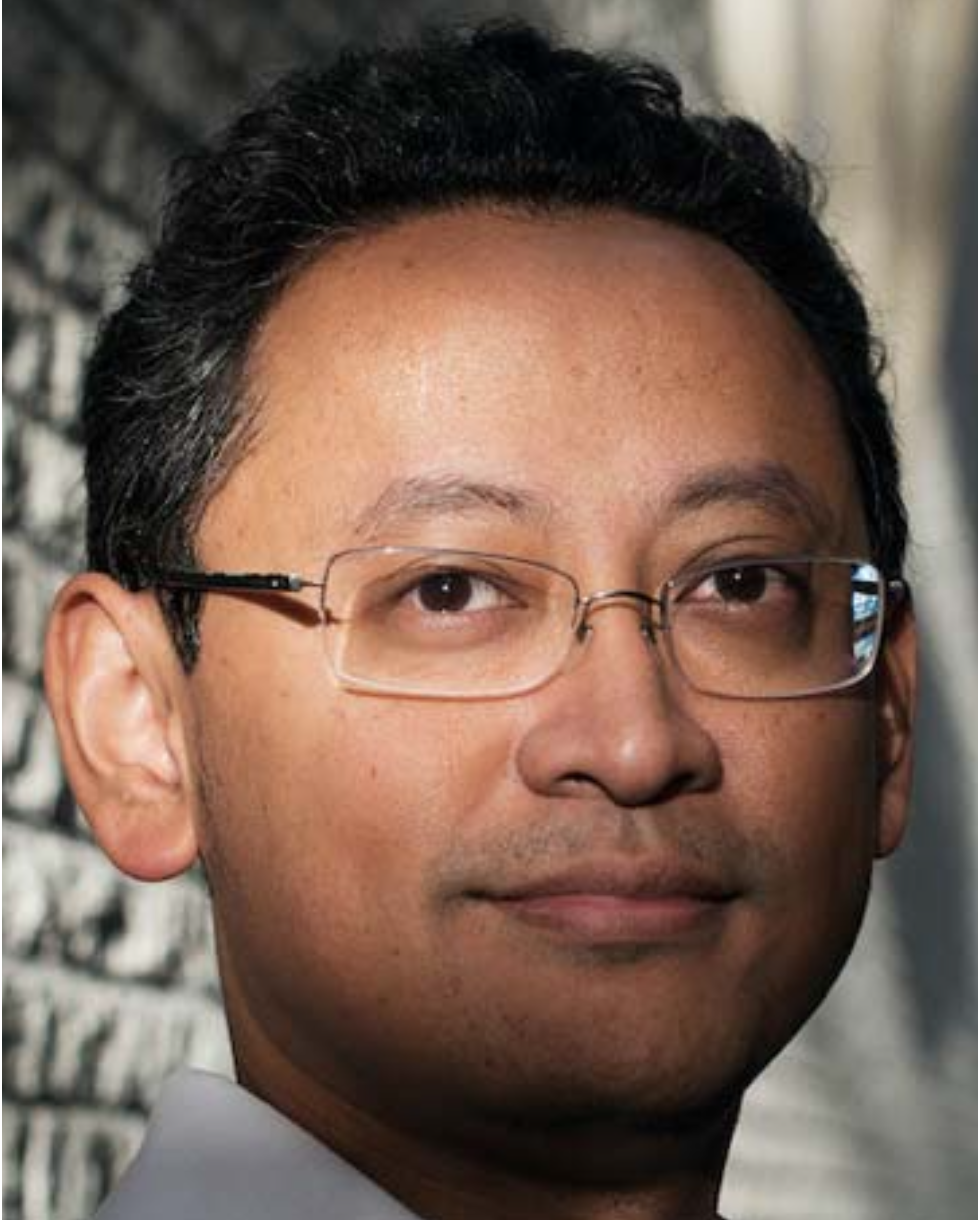}}]{Bayu Jayawardhana} (Senior Member, IEEE)
received the B.Sc. degree in electrical and electron-
ics engineering from Institut Teknologi Bandung,
Bandung, Indonesia, in 2000, the M.Eng. degree in
electrical and electronics engineering from Nanyang
Technological University, Singapore, in 2003, and
the Ph.D. degree in electrical and electronics engi-
neering from Imperial College London, London,
U.K., in 2006. He was with the University of
Bath, Bath, U.K., and with the Manchester Inter-
disciplinary Biocentre, University of Manchester,
Manchester, U.K. He is currently a Full Professor with the Faculty of Science
and Engineering, University of Groningen, Groningen, The Netherlands.
His research interests include analysis of nonlinear systems, systems with
hysteresis, mechatronics, and smart transportation systems. He is a member
of the Conference Editorial Board of the IEEE Control Systems Society,
a Subject Editor of the \textit{International Journal of Robust and Nonlinear Control},
and an Associate Editor of the \textit{European Journal of Control}.
\end{IEEEbiography}

\begin{IEEEbiography}[{\includegraphics[width=1in,height=1.25in,clip,keepaspectratio]{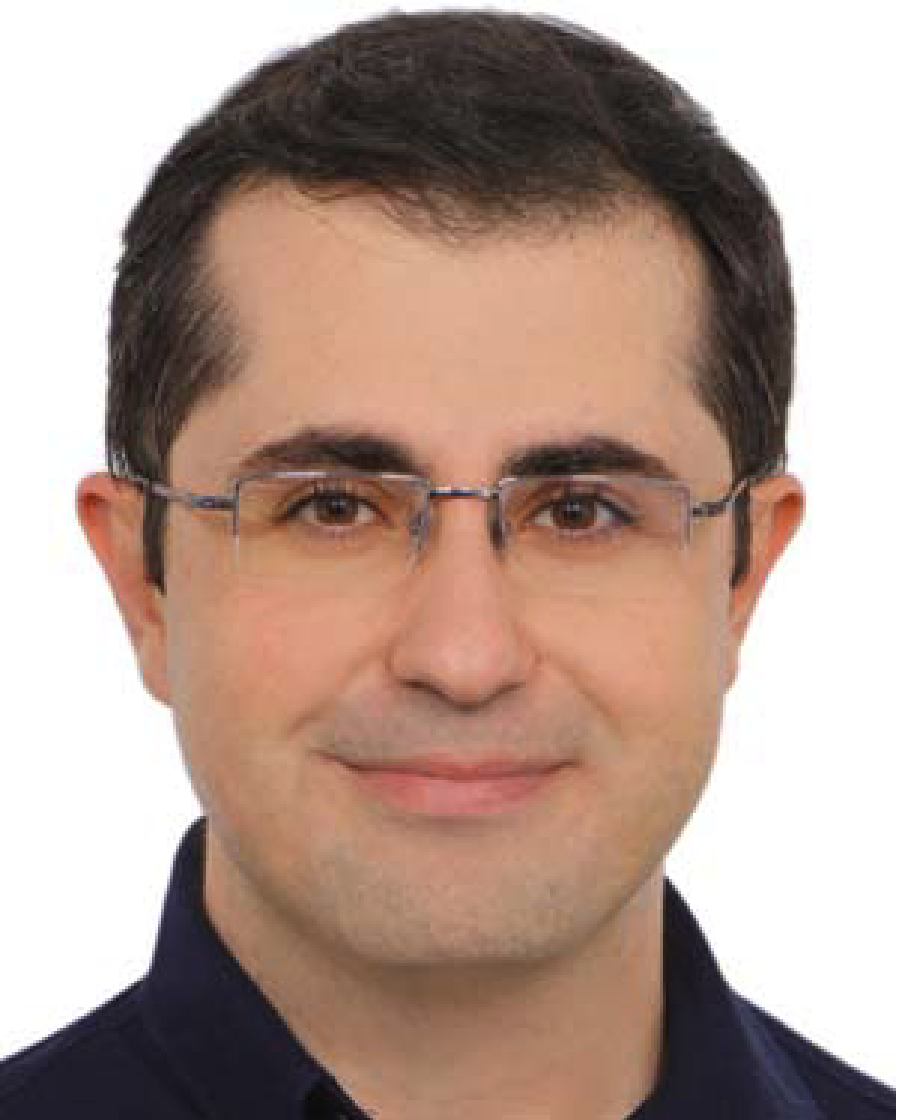}}]{Muhammed O. Sayin} (Member, IEEE) received
the B.S. and M.S. degrees in electrical and
electronics engineering from Bilkent University,
Ankara, Türkiye, in 2013 and 2015, respectively,
and the Ph.D. degree in electrical and computer engineering from the University of Illinois
at Urbana-Champaign, Champaign, IL, USA, in
December 2019.
He is an Assistant Professor with the Department of Electrical and Electronics Engineering,
Bilkent University. He was a Postdoctoral Associate with the Laboratory for Information and Decision Systems, Massachusetts Institute of Technology, Cambridge, MA, USA. His research
interests include developing the theoretical foundation of learning and
autonomy in complex, dynamic, and multiagent systems.
\end{IEEEbiography}

\begin{IEEEbiography}[{\includegraphics[width=1in,height=1.25in,clip,keepaspectratio]{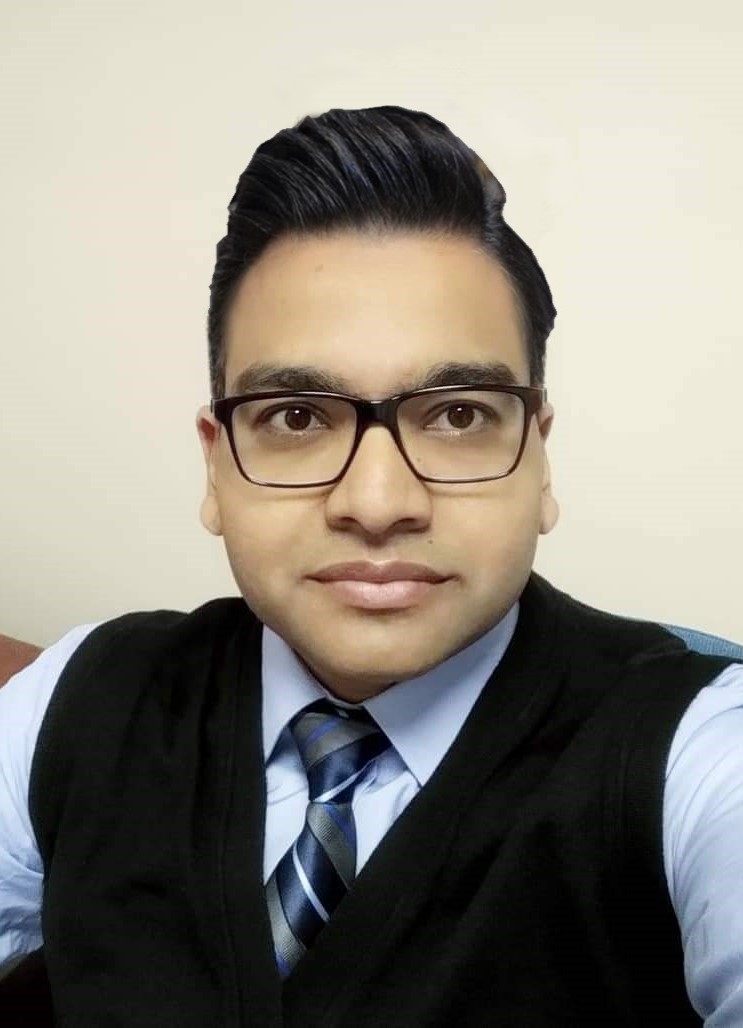}}]{Saeed Ahmed}  is an Assistant Professor of Systems and Control at the University of Groningen, The Netherlands, where he is affiliated with the Engineering and Technology Institute Groningen and the Jan C. Willems Center for Systems and Control. Prior to joining this position, he held postdoctoral positions at the University of Groningen and at the Technical University of Kaiserslautern (now RPTU), Germany. He completed his Ph.D. at Bilkent University, Turkey, in collaboration with Inria, CentraleSupélec, University of Paris-Saclay, France.  His research interests span various topics in systems and control engineering. From a theoretical point of view, he is interested in stability and control, feedback optimization, observer design, and nonlinear and hybrid (switched and impulsive) systems. From an application point of view, he is interested in designing intelligent control algorithms for autonomous vehicles and energy systems.   He received the best presentation award in the Control/Robotics/Communications/Network category at the IEEE Graduate Research Conference 2018 held in Bilkent University, Turkey, the outstanding reviewer award from the European Journal of Control in 2017, and the Investments in Practical Innovations (IPI) Award 2025 from the University of Groningen, The Netherlands. He is an associate editor of Systems and Control Letters and the IEEE Technology Conference Editorial Board (TCEB). He is a member of the IFAC Technical Committees on Networked Systems, Non-linear Control Systems, and Distributed Parameter Systems. He is also the publicity co-chair of the IEEE CSS NextCom committee.
\end{IEEEbiography}

\end{document}